\pdfoutput=1
\documentclass[aps,pre,twocolumn,pre]{revtex4-1}

\usepackage{setspace}
\usepackage{amsmath}
\usepackage{graphicx}
\usepackage{float}
\usepackage{lipsum}
\usepackage{textcomp}
\usepackage{nicefrac}

\usepackage{hyperref}

\begin{document}

%\preprint{}

\title{Receptor-ligand rebinding kinetics in confinement}
%\title{Explicit versus implicit ions generate quantitatively different ligand-binding site dissociation kinetics}
%Authors
\author{Aykut Erba\c{s}}
\affiliation{Department of Materials Science and Engineering, Department of Molecular Biosciences, and Department of Physics and Astronomy, Northwestern University, Evanston, Illinois 60208, USA}
\author{Monica Olvera de la Cruz}
\affiliation{Department of Materials Science and Engineering, Department of Chemistry, Department of Chemical and Biological Engineering, and Department of Physics and Astronomy, Northwestern University, Evanston, Illinois 60208, USA}
\author{ John F. Marko}
\affiliation{Department of Molecular Biosciences, and Department of Physics and Astronomy, Northwestern University, Evanston, Illinois 60208, USA}


%\author{}
%\thanks{}
%\affiliation{}


\date{\today}

\begin{abstract}

Rebinding kinetics of molecular ligands plays a critical role in  biomachinery, from regulatory networks to protein transcription, and is also a key factor for designing drugs and high-precision biosensors.
In this study, we investigate  initial release and rebinding of ligands to their binding sites grafted on a planar surface, a situation commonly observed in single molecule  experiments and which occurs during exocytosis \textit{in vivo}.   Via scaling arguments and  molecular dynamic simulations, we analyze the dependence of non-equilibrium rebinding kinetics  on two intrinsic length scales: average separation distance between the binding sites and dimensions of diffusion volume (e.g., height of the experimental reservoir in which diffusion takes place or average distance between receptor-bearing surfaces). We obtain time-dependent scaling laws for on rates and for the cumulative number of rebinding events (the time integral of on rates) for various regimes. Our analyses reveal that, for diffusion-limited cases, the on rate  decreases via multiple power law regimes prior to the terminal steady-state regime, in which the on rate becomes constant. At intermediate times, at which particle density has not yet become uniform throughout the reservoir, the number of rebindings  exhibits a distinct plateau regime  due to the three dimensional escape process of ligands from their binding sites. The  duration of this regime depends on the average separation distance between  binding sites. Following the three-dimensional diffusive escape process, a one-dimensional diffusive  regime describes on rates.  In the reaction-limited scenario, ligands with higher affinity to their binding sites (e.g., longer residence times) delay the power laws. Our results can be useful  for extracting hidden time scales in experiments where kinetic rates for ligand-receptor interactions are measured in microchannels, as well as for cell signaling via diffusing molecules.

\end{abstract}

\pacs{}
%\keywords{}

\maketitle
%for figure
%\includegraphics[width=9cm,viewport= 0 0 550 800, clip]{Fig.pdf}

\section{Introduction}
The process of diffusion is a simple way of transporting  ligand particles (e.g., proteins, drugs, neurotransmitters, etc.) throughout biological  and synthetic  media~\cite{HowardBerg,Halford:2004iu}. Even though each ligand undergoes simple diffusive motion to target specific or nonspecific binding sites,  ensemble kinetics of these particles can exhibit complex behaviors. These behaviors can be traced back to physiochemical conditions, such as  distribution of binding sites, concentration of ligands in solution, or heterogeneities in the environment.
For instance, biomolecular ligands, such as DNA-binding proteins, can self-regulate their unbinding kinetics via a facilitated dissociation mechanisms dictated by bulk concentration of competing proteins~\cite{Gibb:2014kf,Kamar:2017dd,Hadizadeh:2016hh,Graham:2011cy,Erbas:2018jc,Chen:1jh,Luo:2014ff}. Similarly, spatial distribution of binding sites, such as the fractal dimensions of  a long DNA molecule~\cite{Parsaeian:2013iu}, or surface density of receptors on cell membranes~\cite{GOLDSTEIN:1999dg,Edwards:1997gd,Pageon:2016fk,Sharma:2004} and in flow chambers~\cite{Sorgenfrei:2011eq, Alligrant:2013gv,Guner:2017im}, can influence association and dissociation rates of the ligands.

One way of  probing these molecular reaction rates is to observe the relaxation of a  concentration quench, in which  dissociation of  ligands from their binding sites into a ligand-free solution is monitored to explore the kinetic rates of corresponding analytes~\cite{Myszka:1997tb,Liedberg:1995ws,Schuck:1997wp,Kamar:2017dd,Hadizadeh:2016hh}.   Complete time evolution of this relaxation process depends on factors such as chemical affinity between the binding sites and the ligands,  dimensions of the diffusion volume, and average distance between the binding sites. While the affinity determines the residence time of the ligand on the binding site~\cite{Schwesinger:2000},  the volume available for diffusion can control onset of steady-state  regime at which bulk density of the ligands  becomes uniform throughout the entire diffusion volume.  On the other hand, the spatial distribution of the binding sites can decide how often dissociated ligands revisit binding sites~\cite{Gopalakrishnan:2005gs}.
During the nonsteady state at which average concentration of ligands near the binding sites changes with time, these factors can influence the time dependence of the rebinding kinetics in a nontrivial way.  In turn, these time dependence can effect the kinetic rates prior to the equilibrium.

\begin{figure}[t]
\includegraphics[width=8cm,viewport= 0 10 605 630, clip]{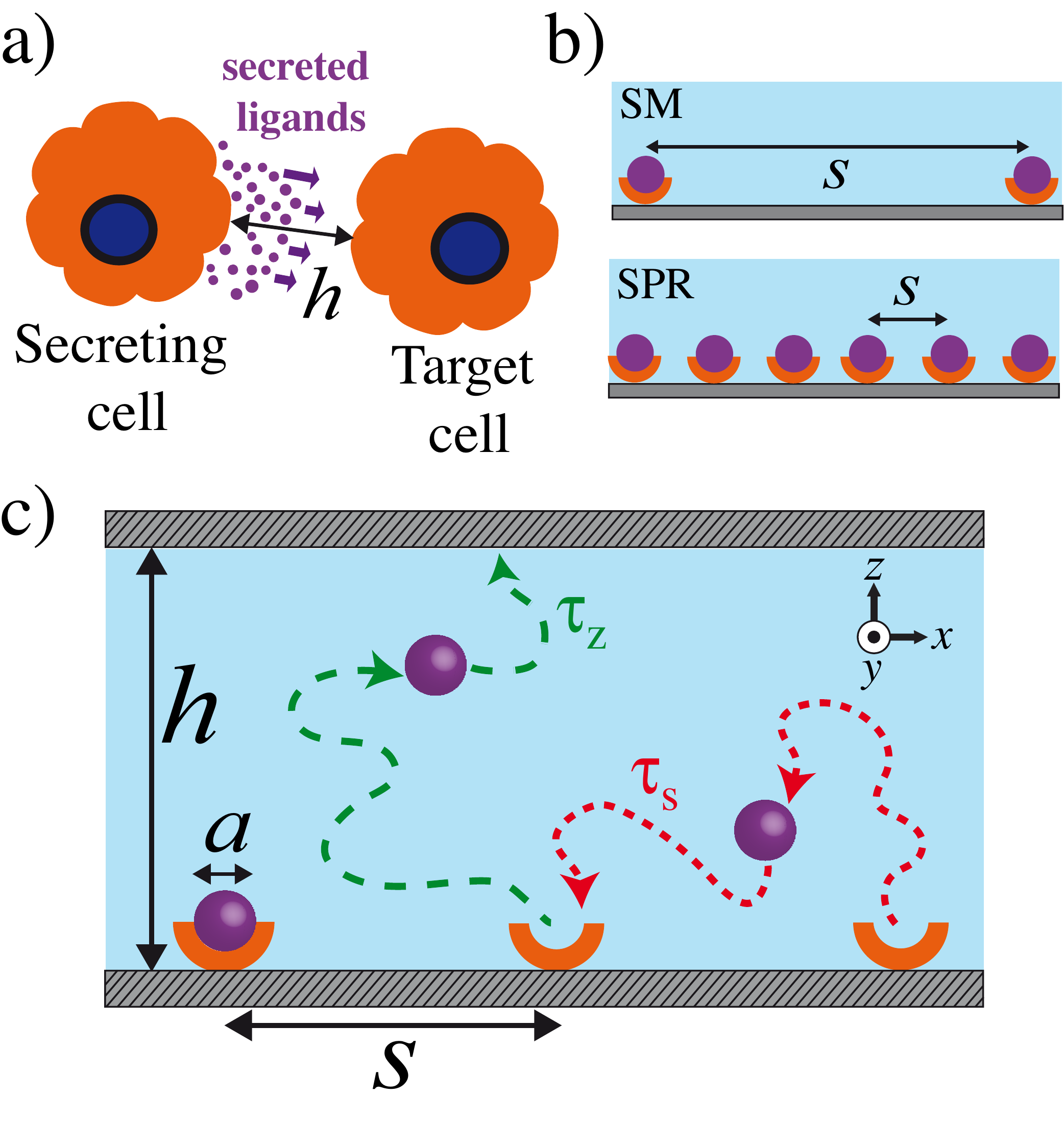}
\caption{a)  Schematics of cell communication via secretion of small ligands into intercellular space of characteristic size of $h$. b) In Single-Molecule (SM) experiments, binding sites (orange) saturated by ligands (purple spheres) are more sparsely distributed compared to SPR experiments. The binding sites are  separated by a distance $s$. c) Illustration of diffusion of ligand particles of size $a$ initially located at their binding sites.  The diffusion volume is confined by two  identical surfaces separated by  a distance $h$. The particles can diffuse to neighboring binding sites within a diffusion time $\tau_{\mathrm{s}}$ and to the confining upper surface within a  diffusion time of  $\tau_z$. Representative trajectories are shown by dashed curves. }
\label{fig:fig1}
\end{figure}

This concentration quench scenerio is common  both \textit{in vivo} and  \textit{in vitro} (Fig.~\ref{fig:fig1}).  In  single-molecule (SM) studies of protein-DNA interactions, short DNA binding sites are sparsely grafted inside a finite-height flow cell~\cite{Kamar:2017dd,Graham:2011cy,Brockman:1999iu}.  The bound proteins may be observed to dissociate into a protein-free solution from their DNA binding sites, allowing measurement of  unbinding kinetics.  Similarly, in Surface Plasmon Resonance (SPR) apparatus, often more densely packed receptors compared to SM experiments are  used to extract kinetic rates. 
On the other hand,   \textit{in vivo}  processes, such as exocytosis and paracrine signaling, in which small molecules are discharged into intercellular space to provide chemical communication between cells, can be examples for  the relaxation of (effective) concentration quenches~\cite{molbiologybook}.
Indeed, due to systemic circulation of ligands  \textit{in vivo} (e.g., time dependent synthesis/ digestion, or phosphorylation/ diphosphorylation of  ligands in cells), a non-steady-state scenario is the dominant situation in biology.

Before the steady state is achieved,  the separation distance between the binding sites (i.e., grafting density of receptors) influences the kinetic rates by regulating rebinding rates~\cite{Gopalakrishnan:2005gs,Lagerholm:1998jw}. Upon the initial dissociation of a ligand from its binding site into solution,  the ligand can return  the same binding site (self-binding) or diffuse to neighboring binding sites (cross-binding) (Fig.~\ref{fig:fig1}c).  In the latter case,  the frequency of rebinding events (i.e., on rate for diffusion limited reactions)  depends on the average distance traveled by the ligand from one binding site to another. Thus, at  time scales comparable to the inter-site diffusion time, the average separation between binding sites becomes a key kinetic parameter.

Experimental studies on ligand-receptor kinetics~\cite{ERICKSON:1987jy,Gopalakrishnan:2005bj} and signal transduction pathways~\cite{Goyette:2017en,Pageon:2016fk,Radhakrishnan:2010fc} have highlighted the critical role  of the spatial placement of  binding sites. 
In additional studies,  in the context of SPR experiments for reservoirs of infinite heights, the effect of  correlated rebinding events on the interpretion of dissociation curves has been brought to attention by using a self-consistent mean-field approximation~\cite{Gopalakrishnan:2005gs,Carroll:2016bp}. However, in those studies,  time scales arising from the diffusion of ligands  from neighboring binding sites have not been distinguished due to  the one-dimensional nature of the analysis.

Inspired by our own and others' experiments, as well as the prevalence of the phenomenon in biological systems, our analysis  focuses on the time evolution of spontaneous dissociation of an ensemble of Brownian particles from their binding sites  into a confined reservoir  (Fig.~\ref{fig:fig1}c). Using scaling arguments and Molecular Dynamics (MD) simulations,  we show that the on rate exhibits two distinct power laws at times longer than initial positional relaxation of the particles but shorter than the time-independent steady-state regime in diffusion-limited reactions. 
We also  derive scaling expressions for the total number of rebinding events experienced by each binding site as a function of time. This quantity can be related to the time-integrated fraction of  bound and unbound ligands in experiments~\cite{Sorgenfrei:2011eq,Kamar:2017dd}.
Our results indicate that the total number of rebinding events exhibits an unexpected plateau behavior at times much earlier than the onset of the steady state. This plateau regime is  terminated by a threshold time scale, which increases with the 4$th$ power of the separation distance. Interestingly, this threshold time scale cannot be detected easily  in the on rate measurements.

Our scaling expressions were compared to  MD simulations  of  ligands modeled as Brownian particles interacting with their binding sites. A detailed analysis of  the simulation trajectories lead to excellent agreements with our scaling predictions. The results that we present here can be applied to new single-molecule studies to determine overlooked time scales involved in binding kinetics  and  can contribute to our understanding of fundamental principles of the kinetic processes in  biological media.

The paper is organized as follows. In Section~\ref{sec:theory}, we  describe scaling arguments and threshold time scales resulted from our scaling calculations. In Section~\ref{sec:MD}, we compare our scaling predictions to  MD simulations.  We then discuss our results and possible indications for single-molecule studies and biological systems, together with some suggestions for possible experiments in the Discussion section.

%%%%%%%%%%%%%%%%%%%%%%%%%%%%%%%%%%%%%%%%%%%%%%%%%%%%
\section{Results}
%%%%%%%%%%%%%%%%%%%%%%%%%%%%%%%%%%%%%%%%%%%%%%%%%%%%%

\subsection{Scaling analysis for ligands diffusing in vertical confinement}
\label{sec:theory}
Consider  $n_0$ identical  particles of size $a$ initially (i.e., at $t=0$) residing on $n_0$ identical binding sites located on a planar surface  at $z=0$ (Fig.~\ref{fig:fig1}c).  A second surface  at $z=h$ confines the reservoir in the vertical (i.e., $\hat{z}$) direction. The size of a binding site is $a$, and the  average separation distance between two binding sites is $s$.    At $t=0$, all particles  are released and begin to diffuse  away from their binding sites into a particle-free reservoir (Fig.~\ref{fig:fig1}c).  This assumption ignores the finite residence times of ligands on their binding sites and will be discussed further in the following sections.

After the initial release of the ligand particles from their binding sites, each particle revisits its own binding site as well as other binding sites multiple times. The on rate,  $k_{\mathrm{on}}$ (proportional to the local concentration of ligands in diffusion limited reactions), and total number of revisits experienced by each binding site, $\mathcal{N}_{\mathrm{coll}}$,  reach their  equilibrium values  once rebinding events become independent of time (i.e., when the ligand concentration in the reservoir becomes uniform). At  intermediate times, during which particle concentration in the reservoir is not uniform, various regimes can arise depending on the separation distance $s$ or the height of the reservoir $h$. 

The time-dependent expressions for $k_{\mathrm{on}}$ and $\mathcal{N}_{\mathrm{coll}}$ prior to the steady state  can be related to the length scales of the system on a scaling level after  making a set of simplifying assumptions. First, we assume that each particle diffuses with a position independent diffusion coefficient, $D$, without hydrodynamic interactions. We also assume that the particles interact with each other, the binding sites,  and surfaces via short-ranged interactions (i.e., interaction range is comparable to the particle size).  This approximation is appropriate for physiological salt concentrations, for which electrostatic interactions are short-ranged.  We also ignore all prefactors on the order of unity.
\begin{figure}[h]
\centering
\includegraphics[width=8cm,viewport= 0 10 550 800, clip]{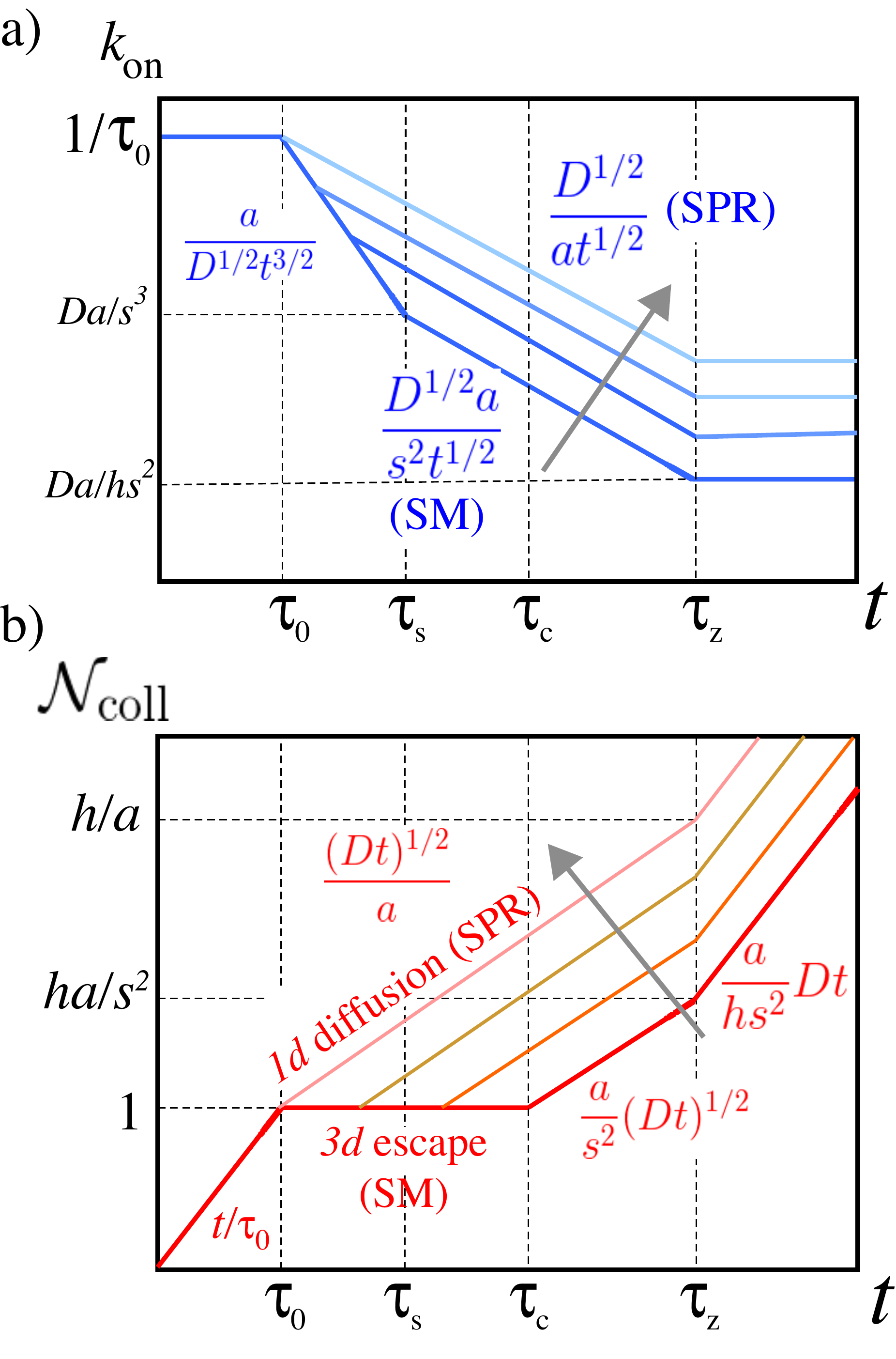}
\caption{Results of scaling arguments for a) the on rates $k_{\mathrm{on}}$ and b) the total number of rebinding events  $\mathcal{N}_{\mathrm{coll}}$ as a function of time in a log-log scale. Arrows indicate the directions of decreasing separation distance between two binding sites (i.e., $s \rightarrow a$).  SM and SPR indicate the regimes related to Single-Molecule and Surface Plasmon Resonance experiments, respectively.  The threshold time scales refer to onset for darker lines. See also Table~\ref{table:table1} for the definition of the threshold times.} 
\label{fig:sketches}
\end{figure}

After the initial dissociation of the ligand particle, particle can explore  a volume $V(t)$ before it revisits any binding site  at time $t$.  If there are  $\omega$ binding sites in $V(t)$, the particle can return any  of $\omega$ binding sites (i.e., $\omega$ is the degeneracy of the binding sites). Thus,  a general scaling ansatz for the on rate can then be written  as
\begin{equation}
k_{\mathrm{on}}(t) \approx  \frac{D a }{V(t)} \omega,
\label{eq:onratescaling}
\end{equation}
Alternatively, Eq.~\ref{eq:onratescaling} can also be interpreted as the inverse of the time that is required for a particle to diffuse through $V(t)/a^3$ discrete  lattice sites if the diffusion time per lattice site is $D/a^2$. Note that for diffusion-limited reactions $k_{\mathrm{on}}(t) \sim c(t)$, where $c(t) \sim V(t)^{-1}$ is the time dependent concentration of ligands.
Note that, for simplicity, in  Eq.~\ref{eq:onratescaling}, we assume that $\omega$ has no explicit  time dependence although this could be added to the  ansatz in Eq.~\ref{eq:onratescaling}. For instance,  for binding sites along a fluctuating chain or for diffusing protein rafts on cell membranes, time dependence can be incorporated without losing the generality of Eq.~\ref{eq:onratescaling}.

The total number of rebinding events detected by each binding site at time $t$ is related to the on rate as
\begin{equation}
\mathcal{N}_{\mathrm{coll}}(t) \approx  \int_{0}^{t} k_{\mathrm{on}}(t') d t'.
\label{eq:Ncoll1}
\end{equation}
\begin{figure*}
\centering
\includegraphics[width=16cm,viewport= 0 20 770 250, clip]{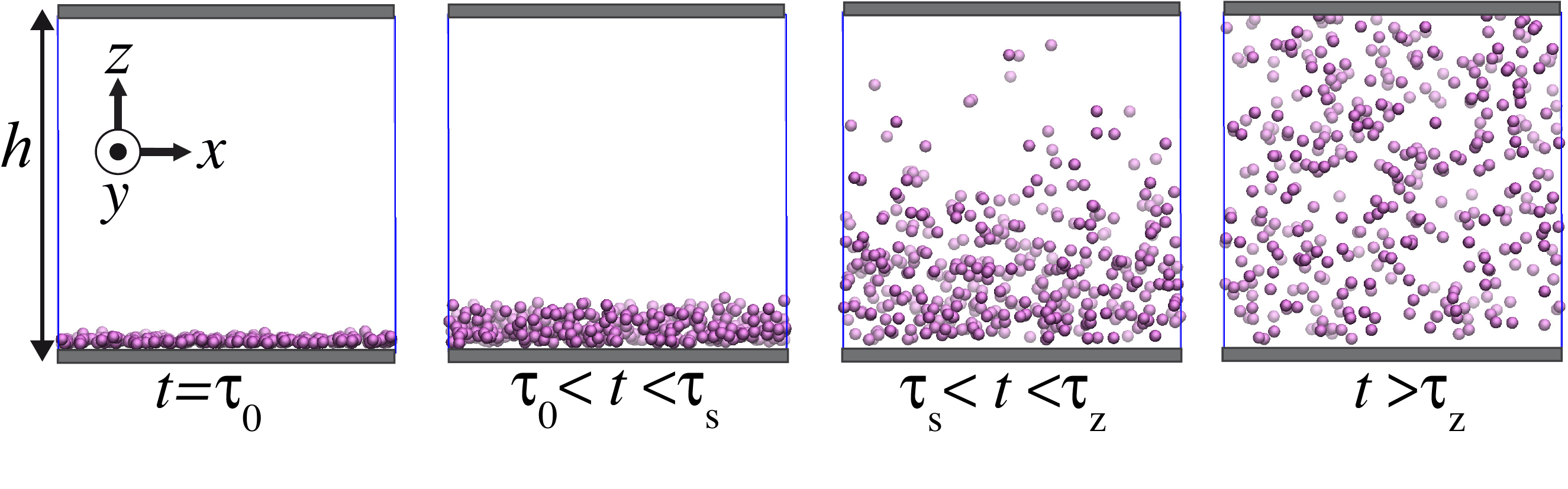}
\caption{Simulation snapshots at various time windows showing the time evolution of the particle concentration throughout a simulation box of  height $h/a= 50$. The separation distance between the binding sites is $s/a=2.5$. The blue lines indicate the  borders of the original simulation box. Periodic boundary conditions are applied only in the $\hat{x}$  and $\hat{y}$ directions, whereas the reservoir is confined in the $\hat{z}$ direction by two identical surfaces. }
\label{fig:snapshots}
\end{figure*}

At the initial times of the diffusion of $n_0$ ligand particles (i.e.,  $t \approx \tau_0 \approx a^2/D$), each particle can undergo  a 3d diffusion process to a distance roughly equivalent to its own size (i.e., self-diffusion distance).  Since, at $ 0<t<\tau_0$,  particles can only collide with their own original binding sites, we have $\omega  \approx 1$, and the interaction volume is $V \approx a^3 \approx (\tau_0 D)^{3/2}$. Thus, according to Eq.~\ref{eq:onratescaling}, the on rate is $k_{\mathrm{on}} = 1/ \tau_0$ and can be considered to be time-independent during $t < \tau_0$  on the scaling level.  From Eq.~\ref{eq:Ncoll1}, a constant on rate leads to a linearly increasing total number of rebinding events as $\mathcal{N}_{\mathrm{coll}} \sim t$ (Fig.~\ref{fig:sketches}). 

For $t>\tau_0$, each particle  can diffuse to a distance $r>a$.  If the separation distance between the binding sites is $s \approx a$, particles can visit any of the nearest binding sites at $t\approx \tau_0$.  If the separation distance  is large (i.e., $s \gg a$), particles  can travel to neighboring sites only after a time  $ \tau_{\mathrm{s}} \approx s^2/D$, at which the average distance traveled by any particle is $s$.  At  $\tau_0 < t < \tau_{\mathrm{s}}$,  individual particles perform 3d diffusion, and thus, the interaction volume is given by $V(t) \approx (D t)^{3/2}$.  Since the volume experienced by particles is $V(t) < s^3$ at $t < \tau_{\mathrm{s}}$, on average one binding site is available per particle in the interaction volume, thus, $\omega \approx 1$. Thus,  using Eq.~\ref{eq:onratescaling}, we can obtain $k_{\mathrm{on}}  \sim t^{-3/2}$. 

Interestingly, at  $\tau_0 < t < \tau_{\mathrm{s}}$, the number of revisits per binding site, $\mathcal{N}_{\mathrm{coll}}$, does not increase  since most particles are on average far away from their own and other binding sites.  On the scaling level, this results in a plateau behavior for the cumulative collision number (i.e., $\mathcal{N}_{\mathrm{coll}} \approx 1$), as illustrated in Fig.~\ref{fig:sketches}. Note that plugging $k_{\mathrm{on}}  \sim t^{-3/2}$ into Eq.~\ref{eq:Ncoll1} leads to a weak explicit time dependence for the $\mathcal{N}_{\mathrm{coll}}$  at  $0<t<\tau_{\mathrm{s}}$ (i.e., $\mathcal{N}_{\mathrm{coll}} \sim 1 +  t^{-1/2}$). The $t^{-1/2}$ dependence indicates that $\mathcal{N}_{\mathrm{coll}}$ stays almost constant during this regime.

At  $t>\tau_{\mathrm{s}}$, the particles can encounter other neighboring binding sites apart from their own, thus, $\omega >1$. The particle density near the bottom surface of the reservoir is nearly uniform, but the overall density is still non-uniform throughout the reservoir. This can be seen in the simulation snapshots shown in Fig.~\ref{fig:snapshots} (we will discuss our simulation results further in the next section).  Only at  a threshold time dictated by the height of the reservoir, $\tau_z \approx h^2/D$, each particle on average reaches the physical limits of the reservoir, and on rate does reach its  steady-state limit (i.e., $k_{\mathrm{on}} \approx D a / h s^2$), as shown in Fig.~\ref{fig:sketches}. At  earlier times, $t <  \tau_z$,  since there are ligand-free regions in the reservoir (Fig.~\ref{fig:snapshots}), $V(t)$ and thus, on rate still must exhibit a time dependence. 

One way of obtaining the time-dependent on rate at $\tau_{\mathrm{s}} < t < \tau_z$  is to consider the diffusion of a single-particle and use $V(t)\approx (Dt)^{3/2}$ and $\omega \approx (Dt/s^2)$ for the number of  binding sites per an area of $(Dt)$. Consequently, Eq.~\ref{eq:onratescaling} leads to  $k_{\mathrm{on}} \approx D^{1/2} a /s^2  t^{1/2} \sim t^{-1/2}$.  This scaling is due to quasi one-dimensional propagation of the particle cloud across the reservoir although each particle undergoes  a 3d diffusion process (Fig.~\ref{fig:snapshots}).  
Alternatively, to obtain the scaling form for on rate, one can consider the overall diffusion of  the particle cloud at $t>\tau_{\mathrm{s}}$ (Fig.~\ref{fig:snapshots}). Then, the total explored volume  scales as $Vt) \sim (D t)^{1/2}$, and  the total number of binding sites in this volume is $\omega  \sim  1/s^2$. Thus,  Eq.\ref{eq:onratescaling} leads to  $k_{\mathrm{on}} \sim t^{-1/2}$.

\begin{table}[t]
\begin{center}
\begin{tabular}{cccccc}
& Scaling & SPR$^{(1)}$  &  SM$^{(2)}$   &  Exocytosis$^{(3)}$  &  Exocytosis$^{(4)}$ \\
 \hline
$\tau_0$       & $a^2/D$        & $10^{-9}$ s    &  $10^{-9}$ s  &  $10^{-9}$ s &  $10^{-9}$ s \\
$\tau_{\mathrm{s}}$      &  $s^2/D$        &   $10^{-6}$ s   & $10^{-2}$ s  &   $10^{-6}$ s &  $10^{4}$ s \\
$\tau_{\mathrm{c}}$      & $s^4/ a^2D$   &  $10^{-4}$ s   &  $10^{4}$ s & $10^{-4}$ s  & $10^0$ s \\
$\tau_z$       & $h^2/D$        &   $10^{2}$ s    &   $10^{6}$ s  &  $10^{-5}$ s & $10^{2}$ s  \\
 \end{tabular}
\caption{The threshold times and their scaling expressions with numerical estimates for various systems. In the estimates,  a ligand of size $a=1\;$nm and a diffusion coefficient of $D=100\;\mu$m$^2$/s are assumed. The estimates are for (1)  $s=10$ nm and $h=10^2\;\mu$m (e.g., SPR case), (2)  $s=1\;\mu$m and $h=10^4\;\mu$m (e.g., SM case),  (3) diffusion of insulin secreted from isolated vesicles into intercellular space with $h=40\;$nm and $s=10\;$nm (determined from  insulin concentration of $\approx 40$ mM in the vesicle \cite{Hutton:1983up}), (4) release of $\approx 10\mu$M of GTP$\gamma$S (i.e., $s=100\;$nm) from a eosinophils-cell vesicle~\cite{Hafez:2003hs}  with an average cell-to-cell distance of $h=100\;\mu$m (i.e., ca. 500 cells per microliter).}
\label{table:table1}
 \end{center}
 \end{table}

The rapid drop of on rate at $t>\tau_0$ with multiple negative exponents has consequences on $\mathcal{N}_{\mathrm{coll}}$.  According to Eq.~\ref{eq:Ncoll1},  the scaling form for $\mathcal{N}_{\mathrm{coll}}$ at  $ t > \tau_{\mathrm{s}}$ can be obtained from the partial  integration of the corresponding  on rate  expressions for appropriate intervals (Fig.~\ref{fig:sketches}a). Thus, at $t>\tau_{\mathrm{s}}$, the total number of rebinding, on the scaling level, is
\begin{equation}\
\mathcal{N}_{\mathrm{coll}} (t) \approx 1 +  \frac{D^{1/2} a}{s^2} t^{1/2}.
\label{eq:eqx}
\end{equation}
We note that inserting $t=\tau_{\mathrm{s}}$ into Eq.~\ref{eq:eqx} leads to  $\mathcal{N}_{\mathrm{coll}} \approx 1$ for any value of $s/a>1$ since the second term on the right hand side of Eq.~\ref{eq:eqx} is smaller than unity. This  indicates that  the plateau regime predicted for $\mathcal{N}_{\mathrm{coll}}$ at $t<\tau_{\mathrm{s}}$  persists even at $t>\tau_{\mathrm{s}}$ (Fig.~\ref{fig:sketches}b).   Only at a later threshold time, $\tau_{\mathrm{c}}>\tau_{\mathrm{s}}$,  the second term of Eq.~\ref{eq:eqx} becomes considerably larger than unity, and so does the  number of revisits, $\mathcal{N}_{\mathrm{coll}}$. The threshold time, $\tau_{\mathrm{c}}$, can be obtained by applying this result on the second term of Eq.~\ref{eq:eqx} (i.e., $D^{1/2}a \tau_{\mathrm{c}}^2/s^2 \approx 1$), which provides an expression for the terminal  time of the plateau regime as
\begin{equation}
\tau_{\mathrm{c}} \approx \frac{s^4}{D a^2}.
\label{eq:tauc}
\end{equation}
At $t > \tau_{\mathrm{c}}$, the total number of revisits per binding site begins to increase above unity. The functional form of this increase at  $\tau_{\mathrm{c}}<t<\tau_z$ can  be obtained  by integrating the on rate (i.e., $k_{\mathrm{on}} \sim t^{-1/2}$) as $\mathcal{N}_{\mathrm{coll}} \sim t^{1/2}$.  This sublinear increase of  $\mathcal{N}_{\mathrm{coll}}$ continues until  the particle density becomes uniform  throughout the entire reservoir at  $t= \tau_z$.  At later times $t> \tau_z$,  diffusion process obeys  Einstein-Smoluchowski kinetics, where the on rate  reaches its time-independent steady-state value, and where  $\mathcal{N}_{\mathrm{coll}}$ increases linearly (Fig.~\ref{fig:sketches}).

To summarize, as also schematically illustrated in Fig.~\ref{fig:sketches}, according to our scaling analysis,  at $t<\tau_0$, the on rate is constant due to self-collisions with the original binding site. At later times,  the on rate decreases as $k_{\mathrm{on}} \sim t^{-3/2}$ until $t<\tau_{\mathrm{s}}$ due to the 3d escape process of particles away from their  binding sites (Fig.~\ref{fig:sketches}a). Once  particles diffuse to distances on the order  of $s$, the particle cloud diffuses in a 1d manner, and the on rate decays with a slower exponent, $k_{\mathrm{on}} \sim t^{-1/2}$. When the particles fill the reservoir uniformly,  a steady-state value of $k_{\mathrm{on}} \sim a/(h s^2)$ takes over. Interestingly,  at the threshold time $\tau_{\mathrm{c}}$, at which we predict a crossover for $\mathcal{N}_{\mathrm{coll}}$, the on rate does not exhibit any alterations and continues to scale as $k_{\mathrm{on}} \sim t^{-1/2}$.

The  regime during which $\mathcal{N}_{\mathrm{coll}}$  is independent of time on the scaling level is smeared out in the limit of  $s \rightarrow a$  as shown in Fig.~\ref{fig:sketches}b. If  $s = a$, the plateau in $\mathcal{N}_{\mathrm{coll}}$ completely disappears, and a scaling  $\mathcal{N}_{\mathrm{coll}} \sim t^{1/2}$ determines the cumulative rebinding events at $\tau_0 <  t < \tau_z$. This indicates that the 3d escape process disappears, and a 1d diffusion-like behavior prevails  after the initial dissociation of ligands. This behavior is  common in SPR experiments, where receptors are often densely grafted.

In the equations below, the scaling expressions for the on rates rescaled by $1/\tau_0 \approx (D/a^2)^{-1}$ and the total number of  revisits are given together with their respective prefactors for corresponding time intervals (see Table~\ref{table:table1}) as 
\begin{equation} 
k_{\mathrm{on}}   \tau_0                    \approx   \left\{ {\begin{array}{*{20}l}
              1                                             \;\;\;\;\;\; \qquad \qquad \qquad  0 <  t < \tau_0 \\
               (\tau_0 / t)^{3/2}                     \;\qquad \qquad    \tau_0 < t < \tau_{\mathrm{s}}  \\
              (a^2 / s^2)  (\tau_0/ t)^{1/2}    \;\;\;\;  \tau_{\mathrm{s}} < t < \tau_z  \\
              a^3 / (h s^2)                            \;\;\qquad  \qquad  t > \tau_z 
\end{array}} \right.
\label{eq:kon_all}
\end{equation}

Similarly, for the total number of revisits

 \begin{equation}
\mathcal{N}_{\mathrm{coll}}  \approx   \left\{ {\begin{array}{*{20}l}
              t / \tau_0                                   \;\;\;\;\;\;\;\quad \qquad \qquad   0 <  t < \tau_0 \\
              1                                               \;\;\;\; \qquad \quad \qquad \qquad  \tau_0 < t < \tau_{\mathrm{c}}  \\
              (a^2/s^2)  (t / \tau_0)^{1/2}        \;\;\; \;\;\;  \tau_{\mathrm{c}} < t < \tau_z \\
              (a^3 / [h s^2]) (t/\tau_0)              \qquad  t > \tau_z 
\end{array}} \right.
\label{eq:Ncoll_all}
\end{equation}

In Table~\ref{table:table1}, we provide  some numerical examples for the above times scales approximately corresponding to SPR and SM experiments, and exocytosis. Note that the scaling expression given in Eqs.~\ref{eq:kon_all} and~\ref{eq:Ncoll_all} can also be obtained by considering the relaxation of a  Gaussian particle distribution in corresponding dimensions (see Appendix section).

In the next section, we will compare our scaling arguments with the coarse-grained MD simulations and investigate the relation between threshold time scales, and the two length scales $s$ and $h$.

%%%%%%%%%%%%%%%%%%%%%%%%%%%%%%%%%%%%%%%%%%%%%%%%%%%%
\subsection{Comparison with molecular dynamics simulations}
\label{sec:MD}

\subsubsection{Description of simulation methodology}
Coarse-grained MD simulations were performed by using the LAMMPS package~\cite{Plimpton:1995wla}. The ligands that are modeled by spherical beads of size $a$ were placed  on a rigid surface with a prescribed seperation distance of $s$. The surface is also composed of the same type of beads as ligands.  The  ligands  interact with each other and the surfaces via a short range Lennard-Jones potential (see Eq.~\ref{eq:wca} in Appendix section) in an implicit background solvent at  constant temperature~\cite{Mccammon}.  A prescribed number of ligands (i.e., $n_0=400-6400$) are allowed to diffuse into a confined reservoir  at $t>0$. The reservoir is periodic in the lateral directions but confined in the vertical direction by a second surface identical to the first one (Fig.~\ref{fig:snapshots}). 

The rescaled height of the simulation box $h/a$ and the rescaled separation distance $s/a$ were separately varied to monitor their effects on time dependencies of $k_{\mathrm{on}}(t)$ and $\mathcal{N}_{\mathrm{coll}}(t)$.  In the extraction  of $k_{\mathrm{on}}$ values, the binding sites are defined as the initial positions of ligands at $t=0$. Any particle  that is found within the collision range of any binding site (i.e., $r_c/a  = 2^{1/6}$) at a given time $t$  is counted  as a bound particle. In our analyses, $k_{\mathrm{on}}(t)$ is defined as the normalized fraction of  binding sites occupied by ligands for diffusion-limited reactions. For reaction limited case, $k_{\mathrm{on}}(t)$ corresponds to raw dissociation data. The values of $\mathcal{N}_{\mathrm{coll}}(t)$ were calculated via Eq.~\ref{eq:Ncoll1}.  All simulations were carried out until the calculated on rates reached their respective steady states (see Appendix section for further simulation details).
\begin{figure}[h]
\centering
\includegraphics[width=7cm,viewport= 0 0 470 800, clip]{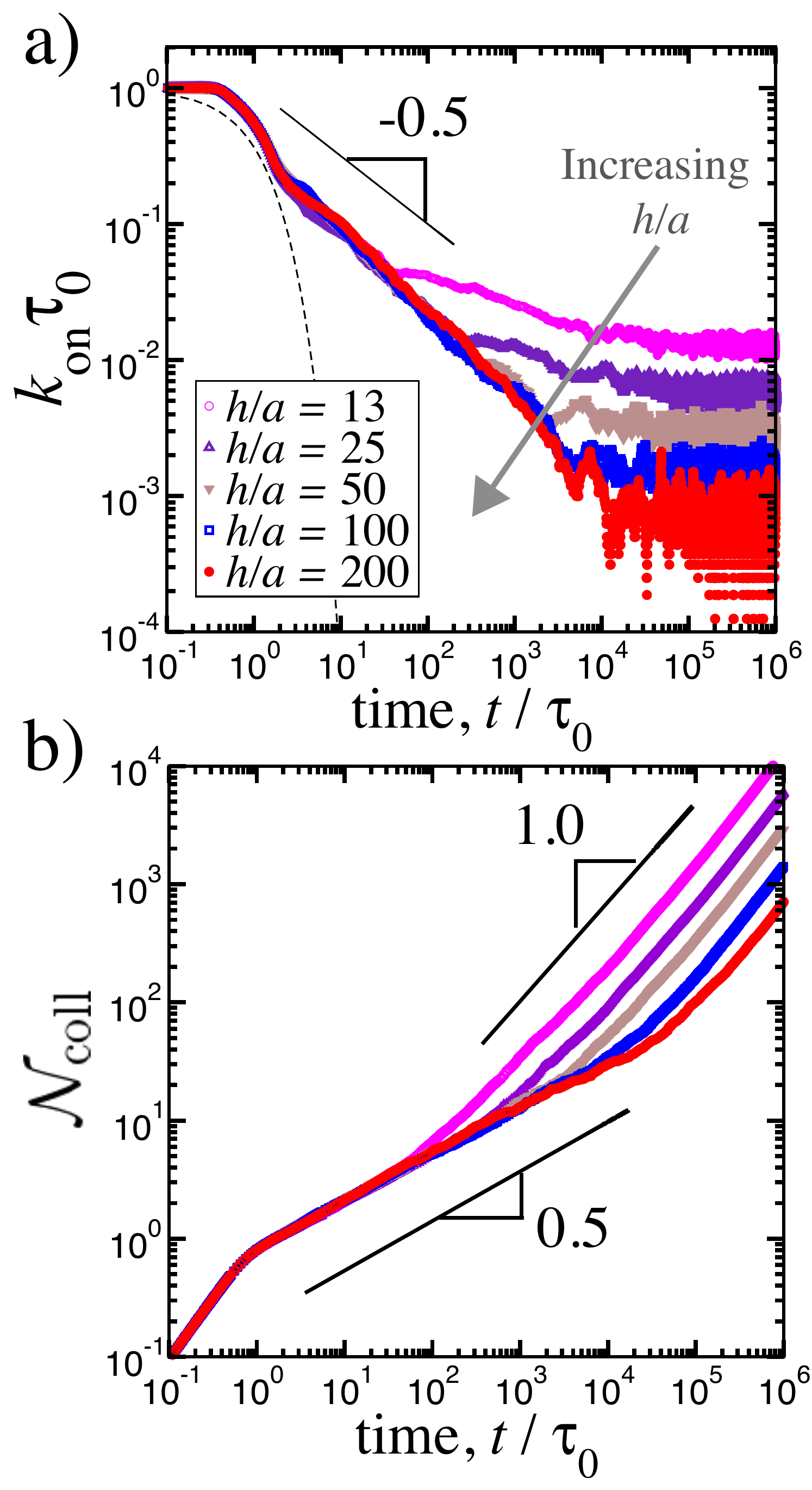}
\caption{a) Rescaled on rates as a function of  the rescaled simulation time for various reservoir heights. The distance between two binding sites is $s/a=2.5$.  A running average over 20 data points is shown for all cases for clarity. b) Total number of revisits  per binding site obtained via Eq.~\ref{eq:Ncoll1} by using non-averaged data series of (a). For all cases, number of ligands particles is $n_0=400$.}
\label{fig:fixeds}
\end{figure}

\subsubsection{Diffusion-limited  kinetics}
\label{sec:diffusion-limited}
We first consider the scenario for which the reactions between the binding sites and ligands are diffusion limited. Thus, the average residence time of the ligand on the binding site is on the order of $\tau_0$, which is the self-diffusion time of a particle in the simulations. We achieved this by  using a purely repulsive WCA potential~\cite{wca:1971} with a  cut-off distance of $r_c/a = 2^{1/6}$ (Eq.~\ref{eq:wca} in the Appendix section). This setup, as we will see, allows us to observe the  regimes predicted in Section~\ref{sec:theory} more clearly. We will further discuss the longer residence times in conjunction with  other time scales in the following  sections.

\paragraph{Qualitative analyses of simulations}
In Fig.~\ref{fig:snapshots}, we present a series of simulation snapshots to demonstrate the diffusion process of $n_0=400$ particles over the time course of the simulations, for $h/a=50$ and $s/a=2.5$. These numbers lead to characteristic times ranging from $\tau_{\mathrm{s}} \approx \tau_0$ to beyond $\tau_z \approx 10^4\tau_0$ for the system shown in Fig.~\ref{fig:snapshots}.
At  short times, $t \approx \tau_0$, the particles are mostly near the reactive (bottom)  surface as can be seen  in Fig.~\ref{fig:snapshots}. As the time progresses, the particle cloud diffuses vertically to fill the empty sections of the box.  At $t < \tau_z$, the particle density near the surface changes with time, and visually the concentration is not uniform in the box. Only for $t> \tau_z$, the particle density becomes uniform, and the initial concentration quench is completely relaxed as illustrated in Fig.~\ref{fig:snapshots}.

\paragraph{Densely placed binding sites in finite-height reservoirs}
To systematically compare our scaling predictions with the simulations, we fixed the separation distance to $s/a=2.5$ and varied the height of the reservoir. Fig.~\ref{fig:fixeds}  shows the data calculated from the particle trajectories  for  the rescaled on rate  $k_{\mathrm{on}} \tau_0$ and the total number of revisits $\mathcal{N}_{\mathrm{coll}}$ as a function of the rescaled simulation time $t / \tau_0$.  
At short times (i.e., $t< \tau_0$), during which particles can diffuse only to a distance of their own size,  $\mathcal{N}_{\mathrm{coll}}$ increases linearly, whereas on rates $k_{\mathrm{on}}$ have no or weak time dependence  in accord with our scaling calculations. In Fig.~\ref{fig:fixeds}a, at approximately $t \approx \tau_0$, we observe a rapid drop in $k_{\mathrm{on}}$, which is  described nominally by an exponential function [$\exp(-t/\tau_0)$]   (dashed curve in Fig.\ref{fig:fixeds}a). However, we should also note that, in the system presented in Fig.~\ref{fig:fixeds}a, $s/a=2.5$, thus,  $\tau_{\mathrm{s}} \approx 6 \tau_0$. Hence, the  decay in the on rate is arguably the beginning of  a power law with an exponent approximately $-\nicefrac{3}{2}$ (Fig.~\ref{fig:sketches}a). 

According to our scaling  analysis,  for small enough separation distances (i.e., $s \approx a$), the on rate obeys a power law (i.e., $k_{\mathrm{on}} \sim t^{-1/2}$) at the intermediate times, $\tau_{\mathrm{s}} <t<\tau_z$ (see Fig.~\ref{fig:sketches}a). In Fig.~\ref{fig:fixeds}a, a slope of $-0.56 \pm 0.04$  describes the decay of the on rates  in accord with our scaling prediction. We have also tested larger systems with $n_0=1600$ and $n_0=6400$ particles and obtain similar exponents. At longer times (i.e.,  $t>\tau_z)$, the on rates  in Fig.~\ref{fig:fixeds}a reach their respective steady-state values  at $t=\tau_z = h^2/D$, which depends on the equilibrium concentration of the ligands  (i.e., $k_{\mathrm{on}}  \sim 1/h s^2$). That is, for a fixed  $s/a$, increasing the height  $h/a$, decreases the concentration and thus the steady-state values of the on rates as seen in Fig.~\ref{fig:fixeds}a. 

As for the total number of revisits, $\mathcal{N}_{\mathrm{coll}}$, in Fig.~\ref{fig:fixeds}b,  the simulation results for densely-packed binding sites show a  power law  dependence on time as $\mathcal{N}_{\mathrm{coll}} \sim t^{0.44 \pm 0.05}$ at the intermediate times in accord with the prediction (Fig.~\ref{fig:sketches}b). Once this regime ends, a subsequent terminal linear regime, in which $\mathcal{N}_{\mathrm{coll}} \sim t^{1.0}$, manifests itself in Fig.~\ref{fig:fixeds}b.  According to our scaling analysis (Eq.~\ref{eq:Ncoll_all}), the onset of this long-time linear regime is set by $\tau_z$.  Thus, increasing the height of the reservoir $h$, only shifts the onset to later times (Fig.~\ref{fig:fixeds}b).  Note that in Fig.~\ref{fig:fixeds}b,  for small values of $h/a$, the exponent is more close to unity since it takes less time to reach a uniform ligand density in smaller simulation boxes. 

\begin{figure}[h]
\centering
\includegraphics[width=7cm,viewport=  0 0  500 830, clip]{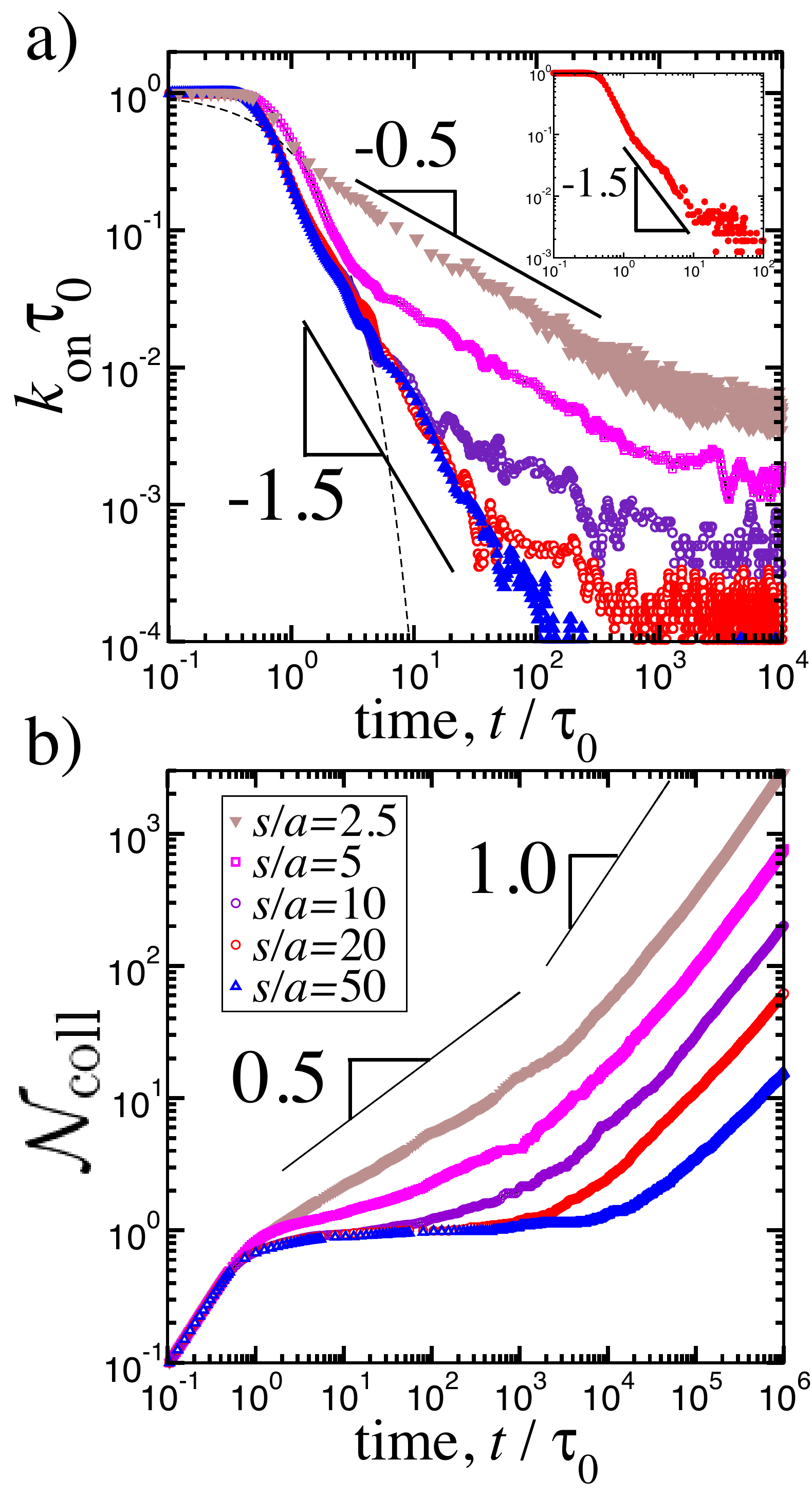}
\caption{a) On rates rescaled by the  unit diffusion time $\tau_0$ as a function of rescaled  time for various separation distances between binding sites for a fixed box height  of $h/a=50$. Each data set is averaged over 3 to 5 separate simulations of the systems containing  $n_0=400-6400$ particles.   For clarity, running averages over 10 points are shown. Inset shows the on rate for a system composed of $n_0=1600$ particles  for $s/a=10$ and $h/a=1000$  b) The total number of rebinding events obtain via Eq.~\ref{eq:Ncoll1} for  $n_0=400$-particle systems by using non-averaged data sets. }
\label{fig:fixedh}
\end{figure}
\paragraph{Effect of separation distance on rebinding kinetics}
As discussed in  Section~\ref{sec:theory},  prior to the steady state,  diffusion time between binding sites significantly affects the apparent dissociation kinetics of ligands. To study this phenomenon, we ran simulations with various values of the separation distance ranging from $s/a=2.5$ to $s/a=50$ for a fixed height of  $h/a=50$ (Fig.~\ref{fig:fixedh}). While  the short-time  kinetic behaviors in Fig.~\ref{fig:fixedh} are similar to those in Fig.~\ref{fig:fixeds} regardless of the surface separation, the long-time  behavior  exhibits various regimes  depending on the separation distance $s/a$ in the simulations.

For $s/a \approx 1$, as discussed earlier, a slope close to $-\nicefrac{1}{2}$ can describe the decay of the on rates  before the time-independent steady state (Fig.~\ref{fig:fixedh}a). For $s/a \gg 1$, this slope is  replaced by a stronger decay  $k_{\mathrm{on}} \sim t^{-1.46 \pm 0.13}$ at  $t>\tau_0$ (Fig.~\ref{fig:fixedh}a).  In our scaling analysis, the exponent $-\nicefrac{3}{2}$ controls the decay of on rate until  the threshold time scale of  $\tau_{\mathrm{s}}$ sets in (Fig.~\ref{fig:sketches}a). Indeed, in Fig.~\ref{fig:fixedh}a, as $s/a$ is increased, the scaling   $k_{\mathrm{on}} \sim t^{-1.5}$  replaces $k_{\mathrm{on}} \sim t^{-0.5}$ type of behavior gradually. For the intermediate values of $s/a$ (i.e., $s/a=5,10,20,50$ in Fig.~\ref{fig:fixedh}a), this transition can be observed  in accord with the scaling prediction in Fig.~\ref{fig:sketches}a.
In the inset of Fig.~\ref{fig:fixedh}a, we also show a system with $s/a=10$ and $h/a=1000$ for a larger system of $n_0=1600$ particles;  The $-\nicefrac{3}{2}$ exponent is more apparent since the two threshold times, $\tau_z$ and $\tau_{\mathrm{s}}$, are well separated due to the large ratio of $h/s \gg 1$.  

 \paragraph{Emergence of plateau behavior in total rebinding events  for sparsely placed binding sites}
The data in Fig.~\ref{fig:fixedh}b  shows the distinct behavior of $\mathcal{N}_{\mathrm{coll}}$ for $s/a \gg 1$ as compared to the cases, where binding sites are closer to each other (Fig~\ref{fig:fixeds}b). As discussed earlier in Fig.~\ref{fig:fixeds}b, for  $s/a \approx 1$, a slope around $\mathcal{N}_{\mathrm{coll}} \sim t^{0.44}$ is dominant  at $t<\tau_z$. However, for $s/a \gg 1$,  a plateau regime replaces this behavior at the intermediate times (Fig.~\ref{fig:fixedh}b).  As  $s/a$ is increased, the plateau regime becomes broader by expanding to longer times. This trend is also in agreement with our scaling analyses (Fig.~\ref{fig:sketches}b). 

The plateau regime in $\mathcal{N}_{\mathrm{coll}}$ is followed by an incremental behavior  as seen in Fig.~\ref{fig:fixedh}b. The predicted power law  following the plateau is  $\mathcal{N}_{\mathrm{coll}} \sim t^{1/2}$ for  $\tau_{\mathrm{s}} < t < \tau_z$ (Fig.~\ref{fig:sketches}b). Within the  duration of our simulations,  we observe  a mixture of  slopes instead of a single exponent of  $\nicefrac{1}{2}$.  For instance, for $h/a=50$, the slope that we can extract at long times is smaller than unity but larger than $\nicefrac{1}{2}$ since $\tau_{\mathrm{s}} \approx \tau_z$ (blue triangles in Fig.~\ref{fig:fixedh}b). This is due to the small ratio of the two threshold time scales, $\tau_z / \tau_{\mathrm{c}} = (h a / s^2)^2 \approx 10$, for the data shown in Fig.~\ref{fig:fixedh}b.
\begin{figure}[!h]
\centering
\includegraphics[width=7cm,viewport= 0 0 650 540, clip]{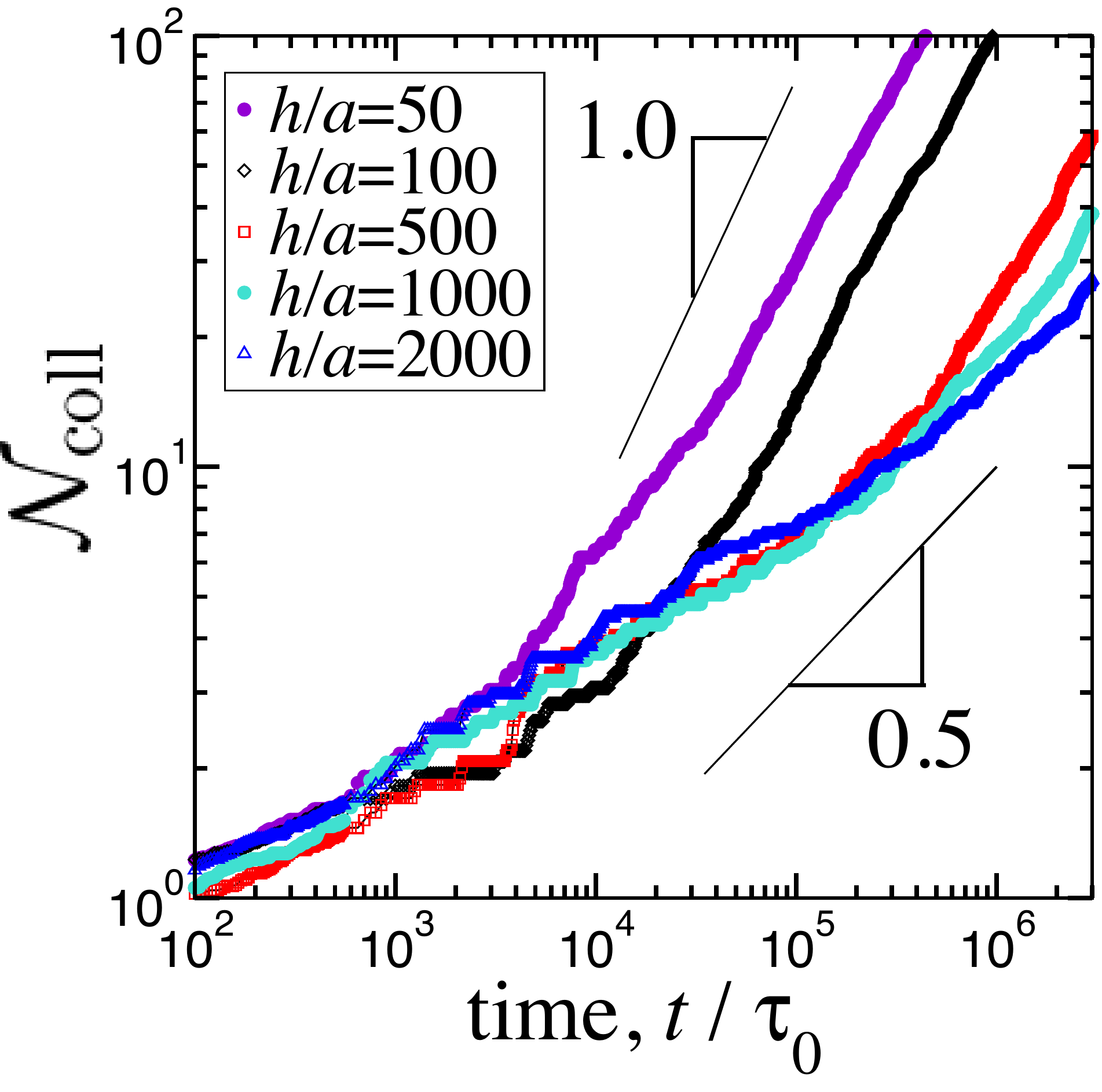}
\caption{The total number of rebinding event as a function of rescaled simulation time for various rescaled heights at a fixed separation distance $s/a=10$. For clarity, only the data for $t > \tau_{\mathrm{s}} \approx 100\tau_0$ is shown. For all cases, number of ligands is $n_0=400$. }
\label{fig:exponent_half}
\end{figure}

To further separate these two time scales, we performed simulations for a fixed $s/a=10$ and for various values of $h/a=50-2000$. The results are shown in Fig.~\ref{fig:exponent_half} for $t>\tau_{\mathrm{s}} \approx 100\tau_0$. For all the data sets in Fig.~\ref{fig:exponent_half}, $\tau_{\mathrm{s}}$ and $\tau_{\mathrm{c}}$ are identical (i.e., equal $s/a$). Thus,  only difference in their kinetic behavior  arises  due to the variations in $h/a$, which in turn determines the duration of  the $\tau_z - \tau_{\mathrm{c}}$ interval. In Fig.~\ref{fig:exponent_half}, ideally, the regime with $\mathcal{N}_{\mathrm{coll}} \sim t^{1/2}$ should be  observable at $\tau_{\mathrm{s}} < t < \tau_z$.  However, we rather observe a weaker increase before a slope of around $\nicefrac{1}{2}$ emerges. We attribute this behavior to the inherent weakness of  scaling analyses since even at $t>\tau_{\mathrm{s}}$, ligands can collide with multiple binding sites frequently enough, particularly for  small separation distances.  These collisions, in turn, can result in a slight increase in  $\mathcal{N}_{\mathrm{coll}}$ similar to that observed in Fig.~\ref{fig:exponent_half}.

At long times $t>\tau_z$,  $\mathcal{N}_{\mathrm{coll}}$ increases with an exponent around unity in the simulations (Figs.~\ref{fig:fixedh}b and~\ref{fig:exponent_half}) in accord with a time-independent  $k_{\mathrm{on}}$. Note that, for   simulations longer than performed here, which are not feasible for computational reasons, we anticipate a convergence to a slope of unity for all of our configurations.
\begin{figure}[h]
\centering
\includegraphics[width=6cm,viewport= 0 0 570 540, clip]{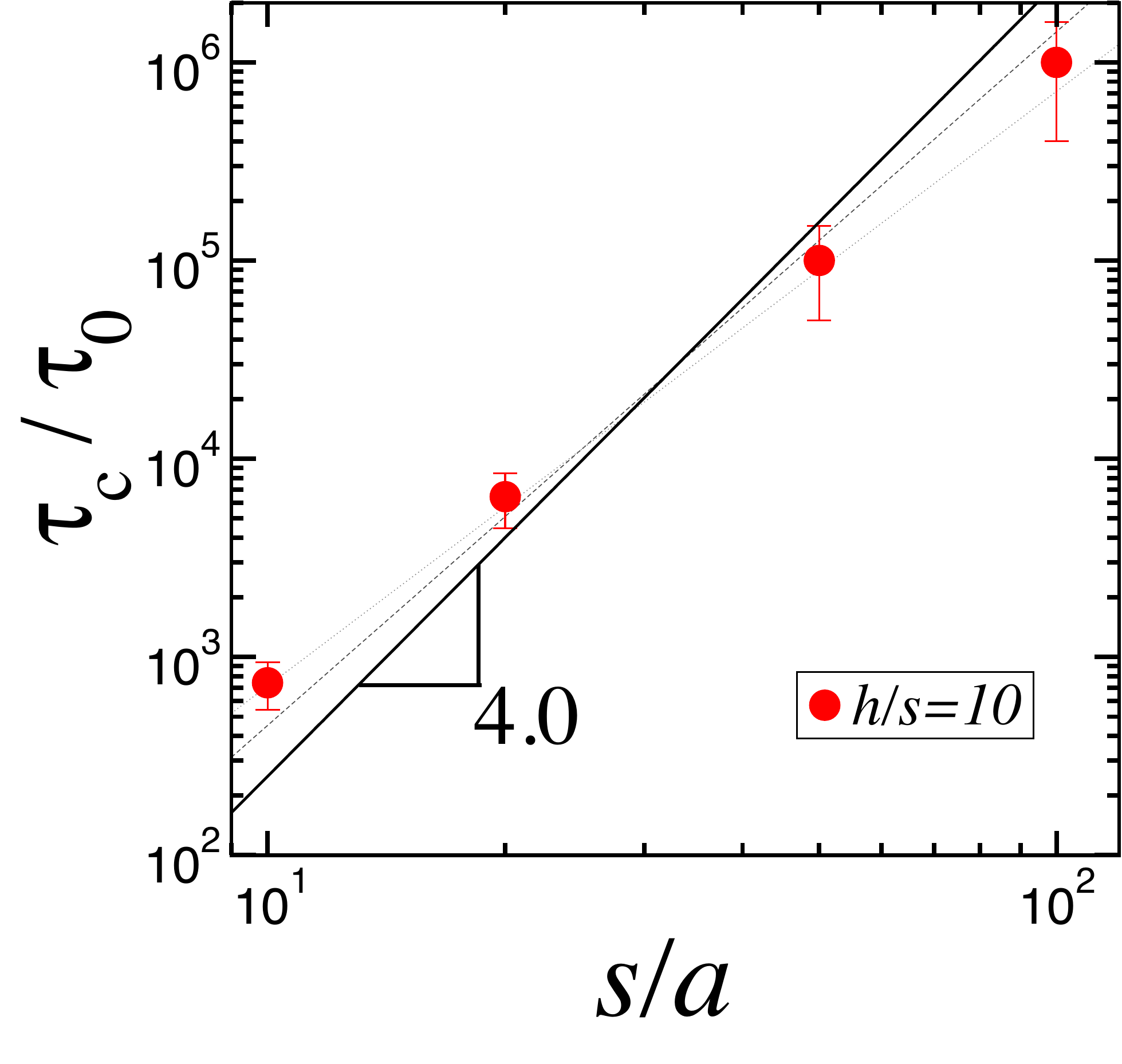}
\caption{Log-log plot of $\tau_{\mathrm{c}}$ extracted from simulations by fitting the respective intervals to a function in the form of $f(t) = 1+ (t / \tau_{\mathrm{c}})^{1/2}$ for $h/s=10$  as a function of the rescaled separation distance.  The thin lines show the slopes of 3.0 and 3.5. For the data sets, $h/a =100,200,500,1000$, and for $s/a=10,20,50,100$.  For all cases, the number of ligands is $n_0=400$. }
\label{fig:tauc_scaling}
\end{figure}

\paragraph{Threshold time for  $\mathcal{N}_{\mathrm{coll}}$ plateau}
We  also performed a separate set of  simulations to specifically identify the scaling dependence of $\tau_{\mathrm{c}}$ on the separation distance (Eq.~\ref{eq:tauc}). We fix the ratio $h/s=10$ and vary the separation distance between  $s/a =10-100$ and  the height between $h/a =100-1000$. We fit  the data encompassing the time interval  $\tau_0 < t < \tau_z$  to a  function in the form of  $f(t) = 1+ (t / \tau_{\mathrm{s}})^{1/2}$  to extract  the threshold time $\tau_{\mathrm{c}}$  at which plateau regimes ends. The results  shown in Fig.~\ref{fig:tauc_scaling} are in close agreement with our scaling prediction; The data can be described by a scaling $\tau_{\mathrm{c}} \sim s^{3.5 \pm 0.5}$. The fact that the exponent extracted from the simulations is smaller than 4 but larger than 2 in Fig.~\ref{fig:tauc_scaling} indicates that the terminal threshold time for the plateau, $\tau_{\mathrm{c}}$, is distinct and well-separated from $\tau_{\mathrm{s}}$.
%
%
%%%%%%%%%%%%%%%%%%%%%%%%%%%%%%%%%%%%%%%%%%%%%%%%%%%%
\subsubsection{Effect of Reaction-limited kinetics}
\label{sec:offrate}
In Section~\ref{sec:diffusion-limited}, we consider the diffusion limited case, where being within the collision range of a binding site is enough to be counted as bound for any ligand. That is $\tau_{\mathrm{off}} \approx \tau_0$. However, most molecular  ligands including DNA binding proteins can have finite lifetimes on the order of minutes to hours~\cite{Kamar:2017dd,Hadizadeh:2016hh}. Long residence times can indeed intervene with the threshold times and regimes predicted by our scaling arguments. 

\begin{figure}[!h]
\centering
\includegraphics[width=8cm,viewport=  0 0  540 800, clip]{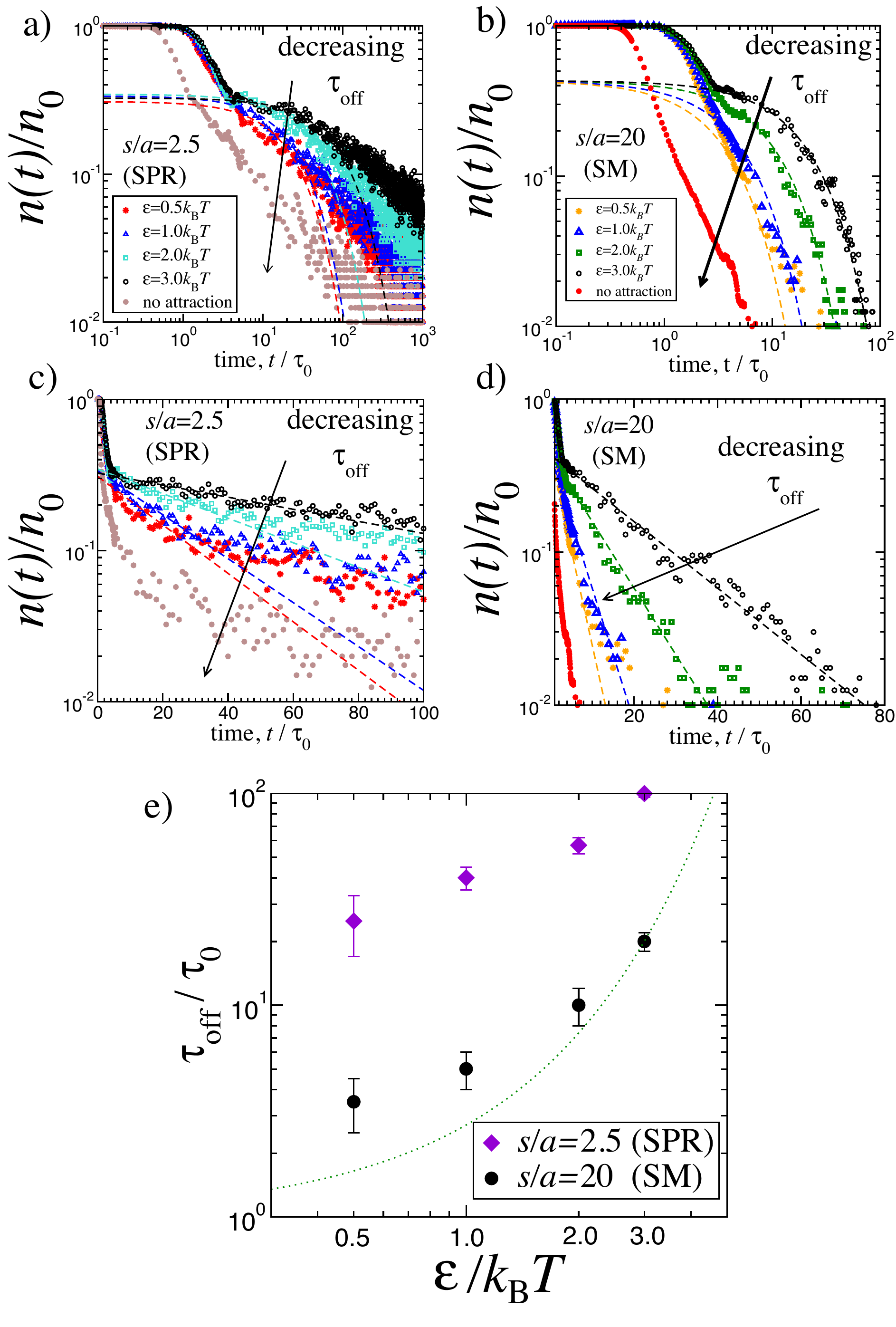}
\caption{Fraction of binding sites occupied by ligands  obtained from the simulations in which a net attraction between binding sites and ligand particles  leads to finite residence times prior to dissociation. The strength of attractions is $\epsilon$ (in the units of $k_{\mathrm{B}} T$) (see Appendix for the pair potential). The brown and red data sets are the same as those in Figs.~\ref{fig:fixeds} and ~\ref{fig:fixedh} with no net attraction. a)  $s/a=2.5$ and b) $s/a=20$, for $h/a=50$. c) and d) log-linear plot of the data sets in (a) and (b). The dashed curves are exponential fits ($\approx \exp(-t/\tau_{\mathrm{off}})$) to the data sets for $t \gtrsim 3\tau_0$. e) The log-log plot of residence times obtained from plots a-d as a function of the attraction strength of the interaction potential for $s/a=2.5$ and $s/a=20$. The dotted curve is $f(x)= \exp(x/k_{\mathrm{B}} T)$.}
\label{fig:attraction}
\end{figure}

To test how the finite residence times can effect the rebinding rates, we ran a separate set of simulations, in which a net attraction is introduced  between the binding sites and the ligands, for two different separation distances, $s/a=2.5$  and $s/a=20$, with $h/a=50$ (Fig.~\ref{fig:attraction}). The attraction was provided  by increasing the cutoff distance  and varying the strength of the interaction potential in the simulations (see Appendix for details).  As a result of this net attraction,  the ligands stay on their binding sites for longer times (i.e., $\tau_{\mathrm{off}} > \tau_0$).  
Importantly, the data presented  in  Fig.~\ref{fig:attraction}  corresponds to the fractions of occupied binding sites since on rate is no more proportional to the concentration in the reaction-limited case.

 In Fig.~\ref{fig:attraction}, at the short times, we observe a rapid drop, regardless of the strength of the attraction. We attribute this common initial behavior to the escape process of the ligands from the attractive potential. After the rapid decay, for high affinities (longer lifetimes), the regimes with either $-\nicefrac{1}{2}$ or $-\nicefrac{3}{2}$ exponents disappear and an exponential function in the form of  $\exp(-t/\tau_{\mathrm{off}})$ can describe the data at the  intermediate times  (dashed curves in Fig.~\ref{fig:attraction}a-b). This can also be seen in the log-linear plots in Fig.~\ref{fig:attraction}c and d.
 As the attraction strength is decreased, the power laws become dominant again as expected from  the diffusion limited cases (Fig.~\ref{fig:attraction}a-d).  
 For longer simulations,  we anticipate that the power laws should be attainable if $s$ and $h$ are large enough.  This can  be seen in Fig.~\ref{fig:attraction}a; after the exponential decay at around $t=100\tau_0$, a slope of $-\nicefrac{1}{2}$  begins to emerge.  We will further discuss the criterion for observing an exponential decay in the Discussion section (Fig.~\ref{fig:phase})

In Fig.~\ref{fig:attraction}e, the residence times $\tau_{\mathrm{off}}$ extracted from the exponential fits are shown for two separation distances, $s/a=2.5$  and $s/a=20$. Even though the attraction strengths between the ligands and binding sites are identical  for two cases (i.e., $\varepsilon = 3,2,1, 0.5k_{\mathrm{B}} T$), the extracted lifetimes are longer for the smaller separation distance (Fig.~\ref{fig:attraction}). This difference  highlights  that the rebinding of ligands from neighboring binding sites  can influence measurements of intrinsic rates. Particularly, for weakly binding ligands, the lifetimes, thus the off rates, are overestimated for the systems in which binding sites are closer.

Overall, our MD simulations support the scaling analyses suggested in Section~\ref{sec:theory} for rebinding rates as well as total rebinding statistics. All regimes and their dependencies on two parameters, $h$ and $s$, are in good agreement with the data extracted from our constant temperature simulations. Below we will  discuss some implications of our results for various \textit{in vivo} and \textit{in vitro} situations.

%%%%%%%%%%%%%%%%%%%%%%%%%%%%%%%%%%%%%%%%%%%%%%%%%%%
%%%%%%%%%%%%%%%%%%%%%%%%%%%%%%%%%%%%%%%%%%%%%%%%%%%
\section{Discussion}
Collective kinetic behavior of diffusing ligands can exhibit novel properties compared to that of a single  ligand. 
In this study, we focus on the nonequilibrium rebinding kinetics of an ensemble of Brownian ligand particles in a confined volume that is initially free of ligands. Our study shows that nonsteady state on rates $k_{\mathrm{on}}(t)$ and total number of revisits detected by each binding site $\mathcal{N}_{\mathrm{coll}}(t)$ depend on the two  time scales imposed by the two intrinsic length scales of the corresponding system. 

The first length scale  is the largest spatial dimension of the diffusion volume. A steady-state kinetic behavior is reached only when the bulk density of diffusers becomes uniform in the corresponding volume. In experimentally typical flow cells,  this length scale corresponds to the height of microchannel. For \textit{in vivo} diffusion of signaling molecules throughout intercellular void, or in  suspensions of  cells or vesicles, this length scale can be related to average distance between receptor-bearing structures (i.e., $h \sim c_{\mathrm{cell}}^{-1/3}$, where concentration of cells is $c_{\mathrm{cell}}$). Once the steady state is reached, the on rate exhibits a time independent behavior  as  $k_{\mathrm{on}} \sim 1/ h s^2$. The rebinding frequency in the steady state is characterized by an Einstein-Smoluchowski limit, which leads to a  linearly increasing $\mathcal{N}_{\mathrm{coll}}$ (Fig.~\ref{fig:sketches}).

The second length scale that we discuss in this work is the average separation distance between two binding sites, $s$, which is inversely proportional to the square root of grafting density of binding sites.  At intermediate times (i.e., before the steady state is established), the on rate  shows one or two  power laws  depending on $s$. For large values of $s$, due to the $3d$ escape process of ligands from their binding sites, the on rate exhibits a $k_{\mathrm{on}} \sim t^{-3/2}$ type of decay after the initial release of the ligand. Once the ligands diffuse to a distance larger than the separation distance $s$, above a threshold time of $\tau_{\mathrm{s}}$, a quasi 1d diffusion process takes over with a smaller decay  exponent of $-\nicefrac{1}{2}$. For densely grafted binding sites (i.e., small $s$), the exponent $-\nicefrac{3}{2}$  is completely smeared out, and the time dependence of the diffusion process is defined by a single  exponent of $-\nicefrac{1}{2}$ at intermediate times  (Fig.~\ref{fig:sketches}a).

We also defined a time-dependent parameter  $\mathcal{N}_{\mathrm{coll}}(t)$  as the time integral of the on rate $k_{\mathrm{on}}(t)$ (more generally, the integral of raw dissociation data) to characterize rebinding kinetics. The parameter $\mathcal{N}_{\mathrm{coll}}$ exhibits a novel plateau behavior on the scaling level  at  intermediate times for sparsely grafted binding sites (Fig.~\ref{fig:sketches}b). The plateau is a result of decreasing probability of finding any ligand near  its binding sites during the $3d$ escape process. This behavior leads to a plateau behavior during which binding sites experience minimal number of collisions with the unbound ligands.  Moreover, due to the integral form of Eq.~\ref{eq:Ncoll1},  $\mathcal{N}_{\mathrm{coll}}$ can be used to invoke the regimes in dissociation measurements otherwise difficult to observe due to relatively noisy statistics  (see Figs.~\ref{fig:fixeds}b and~\ref{fig:fixedh}b).

The plateau  expands to longer times if the binding sites are sparsely distributed since the terminal time  for the plateau scales as  $\tau_{\mathrm{c}} \sim s^4$ (Eq.~\ref{eq:tauc}).  
The termination of the plateau regime is at $\tau_{\mathrm{c}} $ instead of at $\tau_{\mathrm{s}} \sim s^2$. We attribute this behavior to the non-uniform particle distribution near the surface at  $t< \tau_{\mathrm{c}} $; Only after particle the density becomes uniform near the reactive surface, binding sites can experience the incremental collision signals. 
This threshold time  does not manifest itself in the dissociation data  (Figs.~\ref{fig:fixeds}a and ~\ref{fig:fixedh}a) and can be detected only from the cumulative consideration of the dissociation events ( (e.g., Figs.~\ref{fig:fixeds}b and~\ref{fig:fixedh}b)).

 \subsection{Relevance to experiments performed in microfluidic channels}
Experimental studies exploring the kinetics of ligand-receptor interactions, or  single-molecule-based biosensors, are commonly performed in microfluidic channels with well-defined dimensions. We will now discuss some consequences of our study on these experimental systems.
 
 \paragraph{Low grafting density of receptors is essential to extract intrinsic kinetic rates in experiments}
The measurable quantity in kinetic experiments such as  SPR and  fluorescence imaging, is  the population of intact receptor-ligand complexes as a function of time, from which kinetic rates can be obtained.  These apparatus cannot distinguish dissociation and subsequent association of ligands due to their finite-resolution windows. This means that, within the sampling time, a ligand-receptor pair can be broken and reform, possibly with new partners, and thus, contribute to the statistics as an intact complex. This can lead to artificially longer or shorter rates. Our study shows that if  receptors are separated by small distances, the $3d$ escape process is quicker. Thus, rebinding  of ligands  desorbing from nearby receptors can  alter intrinsic rates. We demonstrate this in our simulations (Fig.~\ref{fig:attraction}); Densely placed binding sites lead to longer lifetimes for ligands compared to the case in which binding sites are farther apart.

\paragraph{Association rates can have strong time dependence for weakly binding ligands}
In the kinetic studies of receptor-ligand interactions~\cite{ERICKSON:1987jy} or in modeling signaling pathways~\cite{Aldridge:2006}, time- and concentration-dependent rates in master equations are common practices.  Our study suggests that on rates can have non-trivial time dependence before the steady state is reached for diffusion-limited reactions and weakly binding ligands (e.g., a binding energy on the order of thermal energy).  The time window within which this dependence continues is determined by the dimensions of experimental reservoirs, or average distance between ligand emitting and absorbing surfaces. As an example,  a range of values around $h=10^2-10^4\;\mu$m~\cite{SPRbook,Sugimura:2006co,Kamar:2017dd} leads to $\tau_z \approx  h^2/D  \approx 10^2 - 10^6$ s   if we  assume a diffusion coefficient  of $D = 100\;\mu$m$^2$/s for a ligand of size $a = 1\;$nm (Table~\ref{table:table1}). The estimated values for $\tau_z$  are comparable to the lifetimes of molecular ligands~\cite{Kamar:2017dd}, and the measurement taken earlier may not reflect true on rates, rather quantify an unrelaxed concentration quench.  Note that, in the cases, $\tau_{\mathrm{s}} > \tau_z$,  the regime with a $\nicefrac{1}{2}$ exponent in $\mathcal{N}_{\mathrm{coll}}$ cannot be observed, and a direct transition to the long-time linear regime will be observed (Figs.~\ref{fig:sketches} and~\ref{fig:fixeds}b).

\paragraph{Separation distance brings about its own characteristic time scale}
In  single-molecule fluorescence imaging experiments of  protein-DNA interactions, DNA  binding sites are separated by  distances on the orders of $s \approx 1\;\mu$m~\cite{Kamar:2017dd,Graham:2011cy,Sugimura:2006co}. In SPR experiments, the distance between the surface-grafted receptors is often smaller and can be on the order of $s \approx 10\;$nm~\cite{Nelson:2001wd,SPRbook}. Using ,  the same values for $D$ and $a$ as above, we can obtain some estimates as $\tau_{\mathrm{s}} = 10^{-6} -10^{-2}\;$s and $\tau_{\mathrm{c}} =  10^{-4} -10^{4}\;$s, respectively. While  $\tau_{\mathrm{s}}$, which characterizes the onset of the one-dimensional diffusion regime for on rate, is on the order of tens of miliseconds,  $\tau_{\mathrm{c}}$ can  extend to hours since $\tau_{\mathrm{c}} \sim  s^4$ (Eq.~\ref{eq:tauc}). This wide spectrum of time scales suggests that with adequate design, receptor separations can be used to identify intermingled time scales in an heterogeneous system. 
For instance, biosensors can be prepared with multiple types of receptors (e.g., various nucleic acid sequences), each of which can have a distinct and tractable surface coverage level.  Identification of signals coming from different sets of receptors can allow to interpret  kinetic behavior of certain receptor-ligand pairs, if each separation distance distinctly manifests itself in dissociation kinetics.

\paragraph{Threshold timescales can be used to probe complex systems}
The regimes that we discuss for experimentally measurable on rates and collision numbers  can be used to extract average distance between receptors or receptor bearing surfaces. For instance, the threshold value $\tau_{\mathrm{s}}$ in Eq.~\ref{eq:kon_all} can be utilized to obtain or confirm surface coverage levels of receptors  without any prior knowledge if the decay of the dissociation data is not purely exponential. That is, as we discuss later, $\tau_{\mathrm{s}}$ should be larger than the $\tau_{\mathrm{off}}$.

\paragraph{Prospective experiments to untangle intermingled times scales}
Recently, novel electrochemical-sensor  applications based on the hybridization of a single stranded DNA binding site have been reported~\cite{Sorgenfrei:2011eq,Alligrant:2013gv}. In these experiments, the voltage difference due to hybridization events of the grafted DNA by complementary strands in solution can be measured.  Possibly, in these systems, extremely dilute binding site schemes can be constructed, thus, the time scales we discuss above can be validated experimentally. Indeed, as mentioned above, different nucleic acid strands  can be grafted with varying separation distances, and in principle the resulting signals can be separated since our analysis suggest that each unique separation distance imposes its own terminal threshold times $\tau_{\mathrm{s}}$ and $\tau_{\mathrm{c}}$.

Another experiment setup that  would be interesting could incorporate two SPR surfaces separated by a distance $h$. While  one SPR surface can accommodate receptors saturated by ligands,  opposing surface can have empty receptor sites, hence, create a ``sink'' for the ligands.  Thus,  both rebinding rates and  arrival frequencies can be measured simultaneously. Signals on both surfaces could be compared by systematically varying the density of binding sites, surface separations, etc.
Indeed, this  or similar scenerios can be used to model diffusion of neutransmitters or growth factors \textit{in vivo}~\cite{molbiologybook,Maschi:2017cm}, since rebinding events on the secreting cells can become  slower or faster depending on the number of receptors or their spatial distribution on target cell surfaces~\cite{Pageon:2016fk,Gopalakrishnan:2005bj,Vauquelin:2010bk}. 

 \subsection{Signaling and communication via chemical gradients}
The intercellular void formed by loosely packed cells can percolate to  distances  on the orders of microns~\cite{Handly:2015fs,Francis:1997uv},  which can lead to diffusion times on the orders of minutes. On the other hand,  average (closest) distance between two neighboring cells can be on the orders of 10 nm (e.g., for synaptic cleft). Small molecules, such as cytokines,  secreted from one cell can diffuse throughout these intercellular spaces and provide a chemical signaling system between surrounding cells. This type of communication is controlled by both secretion  and  transport rates~\cite{Ed}.  Indeed, recent studies suggest that spatiotemporal organizations of receptor and ligands can provide diverse signaling responses~\cite{Kholodenko:2006}. In this regard, our result can be used to shed light on some aspects of chemical signaling processes \textit{in vivo} as we will discuss next.

\paragraph{Time-dependent concentration near receptors can provide a feedback mechanism}
Our results suggest that both on rates and total number of rebinding events are sensitive to  time-dependent concentration fluctuations of ligands near  secreting surfaces. According to our analyses,  this time dependence ends when the ligands arrive opposing target surface (e.g., when neurotransmitters diffuse to the receptors of postsynaptic neuron).  This  suggests a feedback mechanism in which the secreting cells can determine  the arrival of the released molecules to the  target cells. This would be possible if the secreting cell bears receptors that are sensitive to the local concentration of the secreted molecule, possibly via time-dependent conformational~\cite{Kobilka:2007tv,Bockenhauer:2011kq} or organizational~\cite{Schmidt:2008,Schutz:2000hd,Simons:2000} changes of membrane components.
In this way,  once the signal molecules reach their target surface, secreting cell can alter the signals depending on the rebinding regime experienced.

\paragraph{Exocytosis can be altered by concentrated vesicles or small openings}
 Our  analysis shows that time-dependent on rates can reach their time-independent regimes faster, and the ligands return their initial position more often,  if  the ligands are closer to each other at the time of the initial  release  (see Figs.~\ref{fig:sketches} and~\ref{fig:fixedh}). In the process of exocytosis of small molecules, vesicles fuse with the plasma membrane and create an opening to release the molecules into intercellular space. One can imagine a scenario, in which, given the concentrations of contents of two vesicles are similar, a ligand released from the vesicle with a larger opening  would return less to the original position (Fig.~\ref{fig:sketches}). If the vesicle opening is small, this would effectively lead to a smaller separation distance, thus, more return would occur.  In fact, the amount of opening can also determine the efficiency of endocystosis (e.g., process of intake of ligand back to vesicle). In exocytosis, secreting vesicles can control the realease rates by changing modes of fusion~\cite{Mellander:2014}. In accord with this concept,  our calculations show that one order of magnitude decrease in the separation distance can increase the return rates by two orders of magnitude (Fig.~\ref{fig:fixedh}a). Similarly,  given the sizes of openings are roughly equivalent for two vesicles, more concentrated vesicle  can lead to more collisions per unit time  with the opposing cell surface, since the number of ligands per unit area is higher during the initial release for the concentrated vesicle  (i.e., $k_{\mathrm{on}}  \sim 1/s^2)$. Similar arguments could be made to explain the observed differences in  exocytosis rates induced by the fusion of multiple vesicles~\cite{Sun:2002ju}.

\begin{figure}[h]
\includegraphics[width=8cm,viewport=  0 0  700 570, clip]{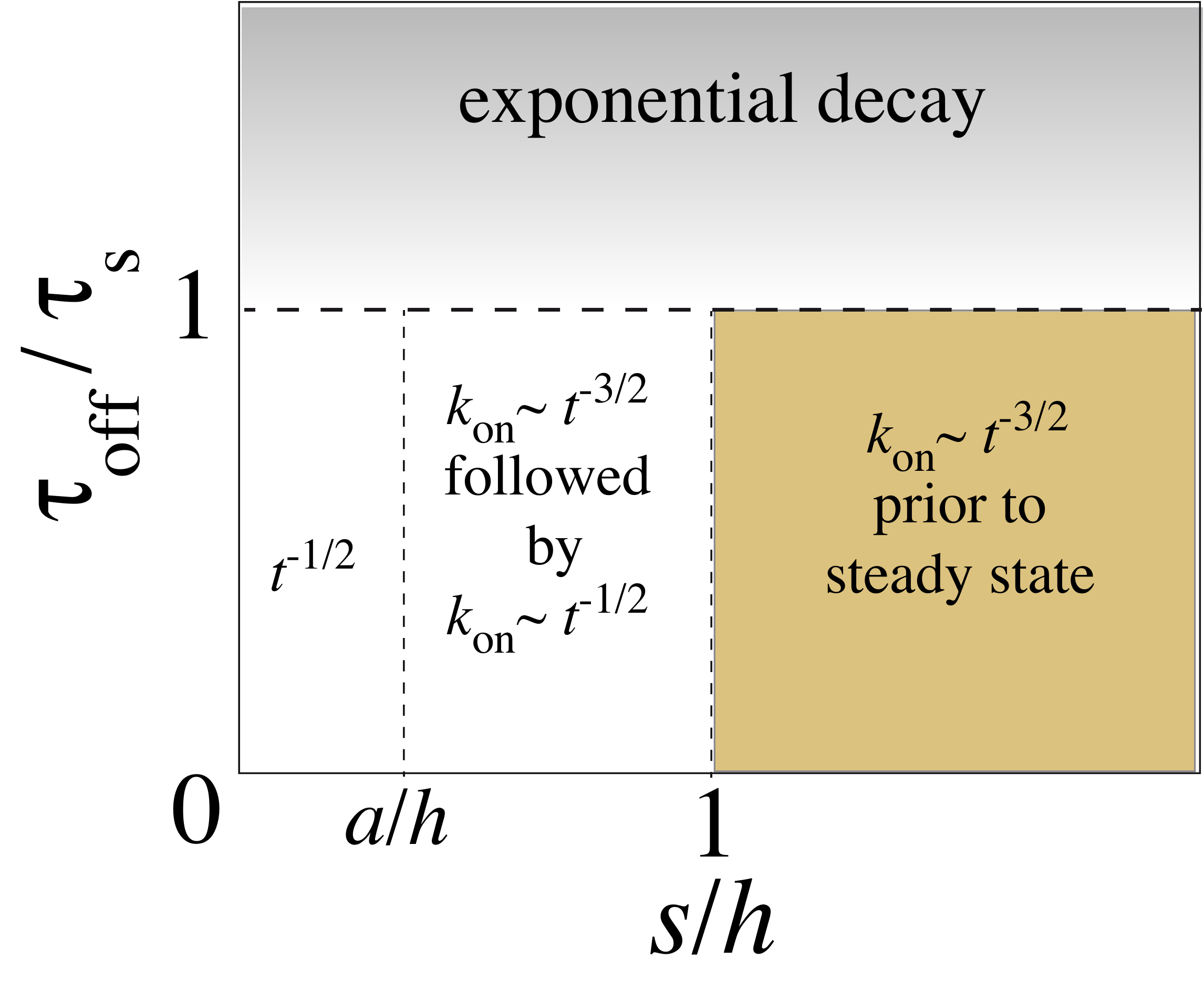}
\caption{Summary of  on rate scaling behaviors predicted for various systems. Along the dashed lines, on rate follows a $-\nicefrac{1}{2}$ exponent.  The residence time of a ligand on its binding site is defined as the inverse off rate, $\tau_{\mathrm{off}} \equiv 1/k_{\mathrm{off}}$. }
\label{fig:phase}
\end{figure}

\subsection{Finite residence time of ligands on binding sites}
In the traditional view of molecular kinetics,  the equilibrium constant of a bimolecular reaction  (e.g., for a protein binding and unbinding its binding site along DNA molecule, or a drug targeting its receptor) is defined as the ratio of off rate $k_{\mathrm{off}} \equiv 1/\tau_{\mathrm{off}}$ and the corresponding on rate.  As we discuss in Section~\ref{sec:offrate},
molecular  ligands can have slow off rates (long residence times) that  can  intervene with the threshold times and regimes predicted by our scaling arguments.  Moreover, these off rates can have strong concentration dependence ~\cite{Kamar:2017dd,Hadizadeh:2016hh}.  Below,  based on recent experimental findings~\cite{Chen:2018hy}, we will briefly discuss some implications of the finite residence times on our results.

\paragraph{Slow off rates can delay power laws} Due to various energetic and entropic components~\cite{Erbas:2018jc,Misra:1994tg}, disassociation process of a ligand from its binding site can be considered as  barrier crossing problem. This rare event manifests itself as a slower decay (compared to a diffusion-limited case) in dissociation curves, which is usually fited by  either exponential or nonexponential curves~\cite{Carroll:2016bp} to extract dissociation rates. For ligands with strong affinity towards their binding sites (i.e., $\tau_{\mathrm{off}} / \tau_{\mathrm{s}} \gg 1$),   this slow decay can occlude the power laws that we discuss, depending on the residence time.
In Fig.~\ref{fig:phase}, we demonstrate the possible effects of the residence times on our calculations with an illustrative diagram in the  dimensions of $s/h$ and $\tau_{\mathrm{off}} / \tau_{\mathrm{s}}$.

In Fig.~\ref{fig:phase}, if the residence time of a ligand is short compared the inter-site diffusion time  (i.e., $\tau_{\mathrm{off}} / \tau_{\mathrm{s}} <1$), the ratio of $s/h$  determines which power law or laws can be dominant at intermediate times. For instance, for $s/h <1$, which is the common scenario in SPR experiments,  both of the decay exponents can be apparent. In other single-molecule experiments,  for which $s/h>1$, $-\nicefrac{3}{2}$ type of exponent can be observable as long as there are enough empty sites for dissociating ligands (i.e., $\tau_{\mathrm{s}}  > \tau_{\mathrm{off}}$).
If the residence  and inter-site diffusion times are comparable (i.e., $\tau_{\mathrm{off}}/ \tau_{\mathrm{s}} \approx 1$), the regime of $k_{\mathrm{on}} \sim t^{-3/2}$ is not observable, and the on rate decays by  $-\nicefrac{1}{2}$ exponent until the steady-state regime.
For a dense array of binding sites, a non-exponential decay can also arise as a result of correlated rebinding events~\cite{Gopalakrishnan:2005gs}.

\paragraph{Time-dependent concentration can induce time-dependent facilitated dissociation}
The recent studies of protein-DNA interactions have shown that off rates, $1/\tau_{\mathrm{off}}$, have a strong dependence on concentration of unbound (free) proteins in solution~\cite{Kamar:2017dd,Graham:2011cy,Chen:1jh}. According to this picture, free ligands in solution can accelarate  dissociation of bound proteins by destabilizing the protein-DNA complex~\cite{Kamar:2017dd,Sing:2014dz}.  Our study shows that, for a concentration quench, concentration of ligands near the binding site changes with time prior to steady state. The time-dependent concentration can  lead to time-dependent facilitated dissociation and shorten the lifetimes of ligands on their binding sites in a time-dependent manner, particularly in the reaction limited scenario. This scenario can be more important when binding sites are too close to each other, since cross rebinding events can induce more facilitated dissociation and further shorten the residence times of the ligands. We do not expect to see this mechanism in our simulations since the facilitated dissociation mechanism usually requires ligands with multivalent nature, which can exhibit partially bound state~\cite{Sing:2014dz}. Future studies can address this mechanism  by considering, for instance, dimeric ligands.

\section*{Author Contributions}

All authors contributed equally to this work. 

\section*{Acknowledgments}

A.E. acknowledges Edward J. Banigan and Ozan S. Sar\i yer for their careful readings of the manuscript, and Reza Vafabakhsh for bringing important literature on synaptic release to our attention.  We acknowledge The Fairchild Foundation for computational support. J.F.M. acknowledges the NIH grants CA193419 and U54DK107980, and  M.O.d.l.C. acknowledges the NSF grant DMR 1611076.

% Uncomment if using bibtex (default)

\subsection*{A. Derivation of on rates via Gaussian distribution}
Here we derive expression for  $k_{\mathrm{on}}$ and $\mathcal{N}_{\mathrm{coll}}$ by using a Gaussian spatial distribution for ligands. Consider  at time $t>0$,  the probability distribution for a set of identical particles in $d$ dimensions evolves from a dirac delta distribution at the origin as
\begin{equation}
P(\vec{r}, t) = \left( \frac{1}{ 2  d D t} \right) ^{d/2} \exp \left( - \frac{ |\vec{r}|^2 }{ 2 \pi d D t} \right),
\label{eq:eq1}
\end{equation}
where $\vec{r}=x_1 \hat{x}_1+x_2 \hat{x}_2 ... + x_d \hat{x}_d$ is the position vector in $d$  dimensions.  The weight of the distributions in Eq.~\ref{eq:eq1} at position $\vec{r}=0$ provides a probability for diffusing particles to revisit the origin 
\begin{equation}
P(0, t) = \left( 2 d D t  \right) ^{-d/2}.
\label{eq:eq2}
\end{equation}
At time $t$, the total number of the revisits can be obtained by integrating Eq.~\ref{eq:eq2}
\begin{eqnarray}
\mathcal{N}_{\mathrm{coll}}(t) & = & \frac{a^d}{\tau_0}  \int_{\tau_0}^{t} P(\vec{0}, t') dt'
             =  \tau_0^{d/2-1} \int_{\tau_0}^{t}  \frac{dt'}{t'^{d/2}}
%             & = & \left( \frac{a^2}{ 2 \pi d D} \right)^{d/2} \int_{\tau_0}^{T}  \frac{dt}{t^{d/2}}\\
%            & = & \left( \frac{2 \pi}{ d D} \right)^{d/2} \frac{1}{1-d/2}  \frac{1}{t^{d/2-1}}
 %            & \sim  &  t^{1-d/2}.
\label{eq:eq3}
\end{eqnarray}

According to Eq.~\ref{eq:eq3}, the returning probability, $P(\vec{0}, t) $, can also be considered as the rate of revisits, $k_{on}$, at the origin $\vec{r}=0$ at a given period $T$. 
\begin{equation}
k_{\mathrm{on}} \equiv  \frac {d \mathcal{N}_{\mathrm{coll}} (t) } { dt }.
\label{eq:eq4}
\end{equation}
After dimensional adjustment, Eq.~\ref{eq:eq4} can also be written as
\begin{equation}
k_{\mathrm{on}} =  a^{d} / \tau_0 P(\vec{0}, t),
\label{eq:eq5}
\end{equation}
From Eq.~\ref{eq:eq5}, the on-rates are
\begin{equation}
k_{\mathrm{on}} = 
\tau_0^{-1}\begin{cases}
   (\tau_0 / t )^{1/2}     & \text{for }  d=1\\
    (\tau_0 / t )               & \text{for }  d=2\\ 
  (\tau_0 / t )^{3 /2}   & \text{for }  d=3 
\end{cases}
\label{eq:eq6}
\end{equation}
\vspace{0.5mm}
If the integral in Eq.~\ref{eq:eq3} can be performed to obtain the expressions for $\mathcal{N}_{\mathrm{coll}}(t)$  as
\begin{equation}
\mathcal{N}_{\mathrm{coll}}(t) \approx
\begin{cases}
    (t/\tau_0)^{1/2}  - 1,     & \text{for }  d=1\\
    \log (t / \tau_0),             & \text{for }  d=2\\ 
     1  - (\tau_0/ t)^{1/2}     & \text{for }  d=3 
\end{cases}
\label{eq:eq7}
\end{equation}
\vspace{0.5mm}
Using Eqs.~\ref{eq:eq6} and~\ref{eq:eq7}, the scaling arguments presented in the main text can be obtained for each step of the diffusion process. Note that even though a $2d$ scenario has been realized for the current problem, it has been show before that  diffusion profile of protein particles that are initially positioned along a one dimensional chain obeys a logarithmic revisit rates~\cite{Parsaeian:2013iu}. 

\section*{Appendices}
\subsection*{B. MD simulation details}
In the simulations, $n_0=400-6400$ binding sites separated by a distance $s$ are placed on a  planar surface composed of square lattice of beads of size $a$ .  To model ligands, $n_0$ beads of size $1\sigma \approx  a$ are placed at contact with binding sites, where $\sigma$ is the unit distance in the simulations.

The  steric interactions between all beads are modeled by a
 truncated and shifted Lennard-Jones (LJ) potential, also known as WCA, 
 \begin{equation}
 V^{\mbox{\tiny LJ}}(r) =\left\{ {\begin{array}{*{20}c}
   4 \varepsilon \left[
(\sigma/ r)^{12} - (\sigma /r)^6 + v_s\right] \;\;r \leq r_c \\
   \;\;0~\quad \qquad\qquad\qquad\qquad\qquad\; r > r_c,
\end{array}} \right.
%\theta(r_c-r).
\label{eq:wca}
\end{equation}
where  $r_c$ is the cutoff distance. A cut-off  distances of  $r_c/\sigma=2^{1/6}, 2.5$ are used with a shift factor $v_s=1/4$ for the interactions between all beads unless otherwise noted. The interaction strength is set to $\varepsilon = 1k_{\mathrm{B}} T$  for all beads, where $k_{\mathrm{B}}$ is the Boltzmann constant, and $T$ is the absolute room temperature. For attractive cases, the  cut-off  distance is set to  $r_c/\sigma=2.5$ and the strength of the potential is varied between $\varepsilon = 0.5 - 3k_{\mathrm{B}} T$

All  MD simulations are run with LAMMPS MD package~\cite{Plimpton:1995wla} at constant volume $V$ and  reduced temperature  $T_r=1.0 $.  Each system is simulated for $10^6$ to $2 \times 10^9$ MD steps. The simulations are run with a time step of $\Delta t=0.005\tau$, where the unit time scale in the simulations is $\tau \approx \tau_0$. The data sampling is performed by recording each $1, \;10, \;10^2, \;10^3$ and  $10^4$ steps  for MD intervals $0-10^2$, $10^2-10^3$, $10^3- 10^4$, $10^4- 10^5$ and $10^5 - 10^8$, respectively.  The monomeric LJ mass is $m=1$ for all beads. The temperature is kept constant by a Langevin thermostat with a thermostat coefficient $\gamma=1.0\tau^{-1}$.
  
The volume of the total simulation box is set to $n_0 (s^2 h) \sigma^3$, where the vertical height is $h / \sigma=12.5-2000$. Periodic boundary conditions are used in the lateral ($\hat{x}$ and $\hat{y}$) directions and  at $z=h$ simulation box is confined by a surface identical to that at $z=0$. VMD is used for the visualizations~\cite{vmd}. 

In the fitting procedures, a weight function inversely proportional to the square of the data point is used. Error bars are not shown if they are smaller than the size of the corresponding data point.

% Uncomment if using bibtex (default)
\bibliography{Erbas_etal}

% Uncomment if using biblatex
% \printbibliography

%\section*{Supplementary Material}

\end{document}